\documentclass[
    reprint,
   amsmath,
    amssymb,
    aps,
    nofootinbib
]{revtex4-2}

\usepackage{amsmath,amssymb,mathtools}
\usepackage{bm}
\usepackage{comment}
\usepackage{orcidlink}

\usepackage{graphicx}
\usepackage{dcolumn}

\usepackage{microtype}

\usepackage{siunitx}

\usepackage{xcolor}

\definecolor{bordo}{RGB}{128,0,32}

\usepackage{hyperref}

\usepackage{natbib}

\newcommand{\dd}{\mathrm{d}}
\newcommand{\Msun}{M_\odot}

\begin{document}

\title{Orbital evolution of asymmetric binaries within accreting environments}

\author{Albert Radulea\orcidlink{0009-0001-9905-8100}${}^{1}$}\email{albert.radulea@ensta.fr}
\author{Marcelo Rubio\orcidlink{0000-0002-7298-3128}$^{2,3,4}$}\email{marcelo.rubio@gssi.it}
\author{Konstantinos Kritos\orcidlink{0000-0002-0212-3472}$^{5}$}\email{kkritos1@jhu.edu}
\author{Andrea Maselli\orcidlink{0000-0001-8515-8525}${}^{2,3}$ }\email{andrea.maselli@gssi.it}
\affiliation{\vspace{0.2cm}
${}^{1}$École Nationale Supérieure de Techniques Avancées (ENSTA), Institut Polytechnique de Paris, Rte de Saclay, 91120 Palaiseau, France\\
${}^{2}$Gran Sasso Science Institute (GSSI), Viale Francesco Crispi 7, I-67100 L’Aquila, Italy\\
${}^{3}$INFN, Laboratori Nazionali del Gran Sasso, I-67100 Assergi, Italy\\
${}^{4}$Grupo de Relatividad y Gravitación, Facultad de Matemática, Astronomía, Física y Computación,
Universidad Nacional de Córdoba, Córdoba PC 5000, Argentina\\
${}^{5}$William H. Miller III Department of Physics and Astronomy, Johns Hopkins University, 3400 North Charles Street, Baltimore, Maryland 21218, USA
}

\begin{abstract}
Extreme mass-ratio inspirals embedded in 
accretion disks provide a natural arena for 
studying the interplay between relativistic orbital 
dynamics and environmental effects. In this work, we develop a 
framework to investigate the secular evolution of 
compact objects repeatedly crossing an accretion 
disk around a supermassive black hole. The orbital 
motion is modeled through Kerr geodesics, 
while disk interactions are encoded through 
effective prescriptions for mass accretion and 
dynamical friction. We find that disk-induced dissipation generically 
drives a two-stage evolution characterized by rapid 
alignment of the orbital plane with the disk, followed 
by slower eccentricity damping. By systematically comparing the dynamics with a purely Keplerian treatment, we show that cumulative relativistic effects produce deviations even at large orbital separations, where the Keplerian approximation would naively be expected to remain accurate. These discrepancies 
grow through repeated disk crossings and become 
increasingly pronounced in more relativistic orbital 
configurations.
We further investigate the impact of the accretion-disk 
model by comparing the Sirko--Goodman and Novikov-Thorne prescriptions. Relativistic disk structures predict systematically 
lower densities and larger scale heights, leading to 
weaker orbital dissipation and slower secular evolution. 
By contrast, the spin of the central black hole has only 
a minor effect on the overall circularization efficiency. Our results demonstrate the importance of consistently 
modeling both relativistic orbital dynamics and disk structure when studying compact objects embedded in AGN disks, and provide a framework for exploring their long-term evolution, as well as a possible connection to quasi-periodic eruptions.
\end{abstract}

\maketitle


\section{Introduction}
\label{Sec-Intro}

Compact objects embedded in gaseous environments provide 
a unique laboratory for studying the interplay between 
orbital dynamics and astrophysical processes \cite{Cardoso:2019rvt,Barausse2014_EnvEffects}. In many 
realistic scenarios, binaries do not evolve in vacuum, 
but interact continuously with the surrounding medium 
through mechanisms such as gas accretion, dynamical 
friction, hydrodynamic drag, and gravitational torques. 
These interactions can modify the orbital evolution over 
long timescales, altering quantities such as the 
eccentricity, inclination, and semi-major axis of the 
system. Understanding such effects is therefore essential 
for constructing realistic evolutionary models of compact 
binaries in dense astrophysical environments.

The relevance of these processes is further enhanced by 
the advent of next-generation gravitational-wave observatories. Future detectors, including the Laser Interferometer Space 
Antenna (LISA), the Einstein Telescope, and Cosmic Explorer, 
will probe compact binaries over a broad range of masses, frequencies, and astrophysical environments \cite{LISA:2022yao,LISA:2022kgy,Broekgaarden:2023rta}. 
In this context, environmental effects are increasingly 
recognized as a potentially important ingredient in the interpretation of future observations, motivating the development 
of realistic models capable of capturing the interplay 
between orbital dynamics and the surrounding medium \cite{Barausse2014_EnvEffects,PhysRevLett.126.101105,Cardoso:2019rou,Zwick2024_NovelEnvPRD}.

Among the environments proposed in the literature, accretion 
disks surrounding supermassive black holes (SMBHs) are 
particularly interesting. Active galactic nuclei (AGNs) 
provide natural reservoirs of gas in which compact objects 
may be captured, migrate, and interact repeatedly with the 
disk material \cite{Barausse2014_EnvEffects,PhysRevLett.126.101105,Cardoso:2019rou,Zwick2024_NovelEnvPRD}. In this context, disk-induced 
dissipation may significantly modify the orbital evolution, potentially affecting the rates, properties, and 
observational signatures of embedded compact-object 
populations.

Extreme mass-ratio inspirals (EMRIs), consisting of 
stellar-mass compact objects orbiting SMBHs, represent one 
of the most promising systems in which to study these 
phenomena \cite{Amaro-Seoane:2012lgq,Barack:2018yly}. Owing to the large separation of scales between 
the primary and secondary masses, EMRIs can undergo prolonged 
interactions with their environment while remaining sensitive 
to relativistic effects associated with the spacetime 
geometry of the central black hole (BH). In AGN disks, the 
evolution of an EMRI is governed by a combination of 
dissipative mechanisms, including mass accretion, dynamical 
friction, hydrodynamic drag, and the gravitational influence 
of the disk itself \cite{BarausseRezzolla2008,ViscousTorqueGN2024}. 
These effects can drive orbital circularization, alignment 
with the disk plane, and long-term migration through the 
gaseous medium.

The efficiency of such processes remains an open question. 
While early studies suggested rapid alignment and circularization 
of embedded compact objects, more recent investigations have 
shown that the outcome depends sensitively on the disk 
properties, orbital configuration, and the nature of the 
disk--object interaction \cite{PhysRevLett.129.241103,Grishin:2023riv,Sedda:2023big,fxnd-r4bs}. 
In particular, the transition from nearly circular to highly eccentric orbits can substantially modify the dominant 
dissipative mechanisms, suppressing migration torques and 
enhancing the role of local dynamical friction.

Despite the growing interest in compact objects embedded in 
AGN disks, most existing studies rely on Keplerian  
descriptions of the orbital motion \cite{10.1093/mnras/staa3004}. Such approximations are 
often well justified at large distances from the SMBH, but they 
become increasingly questionable as the orbit evolves inward 
and relativistic corrections accumulate over many orbital 
periods. Moreover, the interplay between relativistic orbital 
dynamics and disk-induced dissipation remains poorly 
understood, particularly for generic eccentric and 
inclined configurations undergoing repeated disk crossings.

Understanding this interplay is important for several reasons. 
From a theoretical perspective, it is necessary to establish 
the validity of commonly adopted approximations and to quantify 
the impact of relativistic corrections on the secular evolution 
of embedded compact objects. More broadly, these systems have 
been proposed in a variety of astrophysical contexts, including 
AGN-assisted EMRI formation channels and models for 
quasi-periodic eruptions (QPEs), in which repeated disk 
crossings may produce recurring electromagnetic flares \cite{Linial:2023nqs,Franchini:2023bou}.

\begin{figure}
    \centering
    \includegraphics[width=0.5\textwidth]{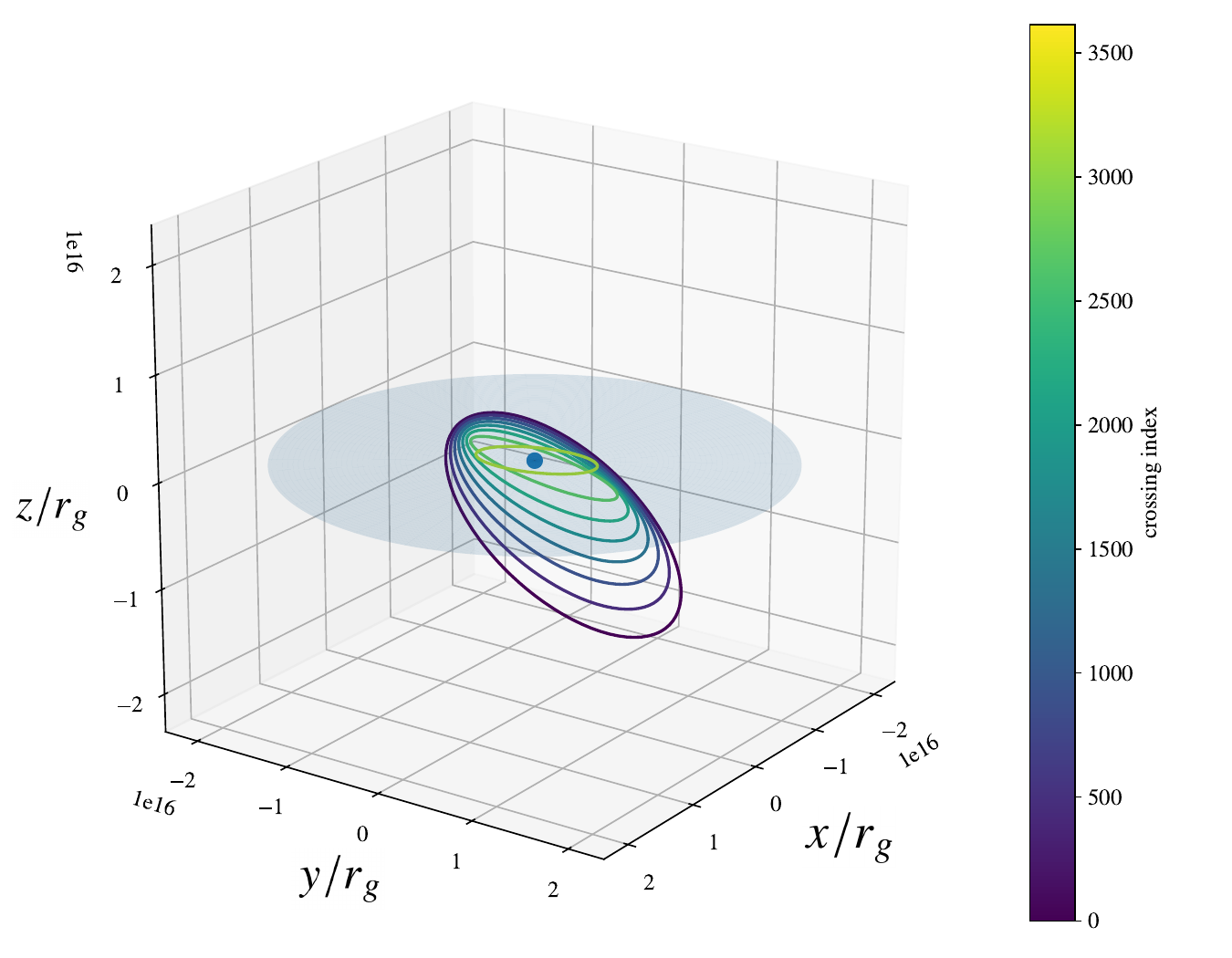}
    \caption{\textit{Orbital dynamics of the binary.} Evolution of the orbital configuration of a compact secondary object inspiraling around a spinning SMBH, in the presence of a thin accretion disk. The colors of the trajectories show the quantity of crossings with the disk. This plot summarizes the main results of our work: repeated disk crossings induce hydrodynamic drag, dynamical friction, and mass accretion effects driving a highly nonlinear secular evolution of the binary. The secondary evolves through an arbitrary number of interactions with the disk, moving on inclined and eccentric orbits. The disk effects align the orbit with the disk plane on timescales much shorter than those associated with gravitational-wave emission, and the eccentricity alternates phases of excitation and damping before the system circularizes (yellow orbit).
}
    \label{Fig1-general-dynamics}
\end{figure}

In this work we develop a relativistic framework to 
study the evolution of compact objects embedded in 
thin accretion disks around supermassive black holes. 
Building upon the recent treatment of Ref.~\cite{Spieksma:2025wex}, where the motion of the 
secondary is described within a mostly Keplerian approximation, 
we model the orbital dynamics through bound Kerr geodesics 
while retaining an effective description of the 
interaction with the surrounding disk. This hybrid approach 
allows us to investigate inclined and eccentric orbits 
undergoing repeated disk crossings, and to assess the 
impact of relativistic effects, black-hole spin, and different 
accretion-disk prescriptions on the long-term evolution of the 
system. In particular, we compare Keplerian and relativistic 
orbital treatments and explore how the predicted evolution 
depends on the underlying disk structure. The main features of the evolution uncovered by our simulations are illustrated in Fig.~\ref{Fig1-general-dynamics}.

The remainder of this paper is organized as follows. 
In Section \ref{Sec-Preliminaries}, we introduce the theoretical 
framework underlying our study, including the 
accretion-disk models and the description of 
disk--secondary interactions. Section \ref{Sec-Numerical-Implementation} presents 
the numerical implementation, with particular 
emphasis on the treatment of Kerr geodesics and 
disk crossings. In Section \ref{Sec-Results}, we discuss the 
orbital evolution of compact objects embedded 
in accretion disks, comparing Keplerian and 
relativistic treatments and assessing the 
impact of different disk prescriptions. Finally, in 
Section \ref{Sec-Discussion-Outlook}, we discuss possible observational 
implications of our results, including their 
connection to quasi-periodic eruptions, as well as a summary of our main conclusions and 
outlines directions for future work.

\section{Preliminaries}
\label{Sec-Preliminaries}

In this section we outline the theoretical framework 
used to describe accretion-disk environments around 
a SMBH of mass $M$ (the primary), 
together with the dynamics of a stellar mass object 
of mass $m$ (the secondary), such that the system 
features a large mass hierarchy, $m/M \ll 1$.
We work in the standard thin-disk regime.

\subsection{Thin accreting environments in a nutshell}

We aim to describe an axisymmetric, stationary, optically thick, 
and geometrically thin disk surrounding a SMBH. Adopting cylindrical 
coordinates $(r,\phi,z)$ centered on the BH, 
we assume that $H/r \ll 1$ throughout the radial extent of interest, where $H$ is the thickness of the disk. Under 
these conditions, the disk structure is governed by mass 
and angular momentum conservation (see for instance \cite{Pringle:1981ds}), namely
\begin{eqnarray}
\dot{M} &=& - 2\pi\, r\, \Sigma(r) \, v_r\,, \label{eq:masscons} \\ 
\dot{M}(\ell-\ell_{\rm \tiny{isco}})&=& 2\pi r^2 W_{r\phi}(r),
\label{eq:angmomintegral}
\end{eqnarray}
where $\Sigma(r)$ is the surface density
\begin{equation}
    \Sigma(r)=\int_{-H/2}^{H/2}\rho(r,z)\,\dd z\ \,,
\end{equation}
and $W_{r\phi}(r)$ is the (vertically integrated) viscous stress,
\begin{equation}
    W_{r\phi}(r)=\int_{-H/2}^{H/2} T_{r\phi}(r)\,\dd z \,.
\end{equation}
Here, $v_r$ denotes the radial velocity and $\ell=r^2\Omega$ 
is the specific angular momentum of the disk. For a Keplerian disk, the angular velocity is $\Omega=\sqrt{M/r^3}$, and the inner edge is located at the innermost stable circular orbit (ISCO), for which $\ell=\ell_{\rm \tiny{isco}}$. 

We model the viscous shear stress following the seminal works by Shakura and Sunyaev \cite{Shakura:1972te,Shakura:1976xk}, which are assumed to be proportional to the total pressure in the disk midplane; i.e.,
\begin{equation}
    T_{r\phi} = -\alpha\left(P_{\rm gas} + P_{\rm rad}\right)\,,
\end{equation}
where $0<\alpha<1$ is a dimensionless viscosity parameter. 
Imposing vertical hydrostatic equilibrium, the effective 
kinematic viscosity can be written as
\begin{equation}
\nu \simeq \frac{\alpha c_s^2}{\Omega}\,,
\label{eq:nu}
\end{equation}
where $c_s = \sqrt{\left(P_{\rm gas} + P_{\rm rad}\right)/\rho}$ 
is the isothermal sound speed. The gas and radiation 
pressure contributions are given by
\begin{equation}
P_{\rm gas} = nk_B T_c\quad \ , \quad P_{\rm rad} = \frac{a}{3}T_c^4\,, 
\end{equation}
with $n$ the particle number density, $T_c$ the 
temperature at the equatorial plane, and $a$ the radiation constant.

\subsubsection*{Disk modeling}

The formalism introduced so far provides a generic 
description of thin accretion disks and is common 
to many of the models considered in the literature 
\cite{Fragile:2024osq,Dihingia:2024tqr,Abramowicz:2011xu}. 
The main differences between these prescriptions arise 
from the treatment of the gravitational field, the role 
of disk self-gravity, and the boundary conditions imposed 
near the central BH. Here we consider 
two disk prescriptions: (i) the classical, non-relativistic 
\textit{Sirko--Goodman} (SG) model, and (ii) the 
relativistic \textit{Penna} model, based on the 
\textit{Novikov--Thorne} (NT) thin-disk framework 
\cite{10.1111/j.1365-2966.2010.17170.x}.

For both prescriptions, we extract the radial profiles 
of the disk density $\rho(r)$, sound speed $c_s(r)$, 
and scale height $H(r)$, which enter directly in the 
computation of dissipative processes affecting the 
secondary motion. 
The two prescriptions predict different density, 
sound-speed, and scale-height profiles, particularly 
in the inner regions of the disk where relativistic 
effects become increasingly important. These profiles 
are tabulated as functions of the cylindrical radius 
$r$ and interpolated numerically at each equatorial 
crossing of the secondary.

The SG model extends the classical Shakura--Sunyaev 
formalism by including the effects of disk self-gravity. 
Such contributions may become comparable to the vertical 
tidal field imposed by the central BH at sufficiently 
large radii. The onset of self-gravity is usually quantified 
through the \textit{Toomre parameter} \cite{Toomre:1964zx}, 
which for a Keplerian disk is given by
\begin{equation}
\mathcal{Q}=\frac{\Omega\, c_s}{\pi G \Sigma(r)}\,.
\end{equation}
When $\mathcal{Q}\lesssim 1$, gravitational instabilities 
lead to the growth of density perturbations, potentially 
resulting in fragmentation and star formation (see 
\cite{kratter2016gravitational} and references therein). 
The central assumption of the SG model is that the feedback 
associated with these instabilities regulates the disk 
towards a marginally stable state, $\mathcal{Q}\simeq 1$. 
This condition provides an additional closure relation for 
the disk structure. As a consequence, the density, temperature, 
pressure, and scale height are determined not only by the 
balance between viscous heating and radiative cooling, but 
also by the requirement that self-gravity remains close to 
the instability threshold \cite{Shakura:1972te,Shakura:1976xk}.

The NT formalism, on the other hand, assumes that the gas 
moves on nearly circular geodesics around the central BH. 
Since the disk is geometrically thin, radial pressure 
gradients and advective energy transport are neglected, and 
the local thermodynamic structure is determined by the balance 
between viscous heating and radiative cooling \cite{Page:1974he}. 
The radial distribution of the emitted flux is then 
uniquely specified by the accretion rate and the 
relativistic properties of the spacetime. In its 
standard form, the model also assumes that the viscous 
torque vanishes at the ISCO, implying that matter 
rapidly plunges into the BH once it reaches 
this radius.

Although the NT model provides the standard theoretical 
description of thin relativistic accretion disks, general-relativistic magneto-hydrodynamic (GRMHD) simulations suggest that magnetic stresses may persist 
inside the ISCO and modify the dissipation profile of 
the inner flow \cite{Agol:1999dn,Krolik:2002ae}. To account 
for these effects, we employ the prescription developed by 
Penna et al. \cite{10.1111/j.1365-2966.2010.17170.x}, which 
calibrates thin-disk models against numerical simulations 
of magnetized accretion flows. In this approach, the NT 
solution remains the underlying description of the disk 
structure, while correction factors derived from GRMHD 
simulations are introduced to model the residual 
transport of angular momentum and energy near the ISCO.

When comparing the two prescriptions, all global 
astrophysical parameters of the accreting system 
are kept fixed. Consequently, any difference in the 
resulting orbital evolution can be attributed 
directly to the distinct density, sound-speed, and 
scale-height profiles predicted by the two disk 
models, rather than to variations in the astrophysical 
setup.

\subsection{Orbital evolution \textit{alla Kepler}}

In this section, we recall the standard Keplerian 
description of disk-crossing dynamics; i.e., modeling 
the interaction between the secondary and the disk. 
In particular, we follow some of the guidelines 
developed in \cite{Spieksma:2025wex}, which prove 
especially useful at large distances from the SMBH. 
In this framework, the secondary follows a Keplerian 
orbit around the central SMBH, while the interaction 
with the gaseous environment is modeled through effective 
dissipative forces. As the orbit approaches the central 
SMBH from generic configurations, relativistic effects 
become increasingly important. For this reason, in this 
work we extend the orbital motion of the secondary to 
Kerr geodesics, while retaining the same effective 
disk-interaction prescriptions in a perturbative sense, 
as shown in the next section.

Following a standard Keplerian description, let us 
consider a secondary BH orbiting around a 
central SMBH. The orbit is characterized by the 
semi-major axis $a$, the eccentricity $e$, and the 
inclination $\iota$ relative to the disk mid-plane. 
The orbital position $\mathbf{r}(t)$ within the orbital 
plane is given by
\begin{equation}\label{rposition-keper}
\mathbf{r}(t) = \frac{a(1-e^2)}{1+e\cos \vartheta}
\left[\cos \vartheta\,\hat{\mathbf{x}} 
+ \sin \vartheta\,\hat{\mathbf{y}}\right]\,,
\end{equation}
where $\vartheta$ denotes the true anomaly, and 
$\hat{\mathbf{x}}$ and $\hat{\mathbf{y}}$ are orthonormal 
unit vectors spanning the orbital plane. We choose 
$\hat{\mathbf{x}}$ to point along the direction of the 
pericentre, while $\hat{\mathbf{y}}=\hat{\mathbf{L}}
\times\hat{\mathbf{x}}$, where $\hat{\mathbf{L}}$ is the 
unit orbital angular-momentum vector. The velocity 
of the secondary relative to the surrounding gas is
\begin{equation}
\mathbf{v}_{\rm rel} = \mathbf{v}_{\rm orb} - \mathbf{v}_{\rm gas}\,,
\end{equation}
with $|\mathbf{v}_{\rm gas}| \simeq \sqrt{GM/r}$ 
for a nearly Keplerian disk.

The total acceleration acting on the secondary can 
be decomposed as \cite{Kley:2012ue}
\begin{equation}
\mathbf{a}_{\rm tot} = -\frac{GM}{r^3}\mathbf{r} 
+ \mathbf{a}_{\rm drag} 
+ \mathbf{a}_{\rm DF} 
+ \mathbf{a}_{\rm disk}\,.
\end{equation}
As expected, in addition to the central gravitational 
force, three disk-induced effects contribute as well: 
hydrodynamic drag due to motion through the gas; 
dynamical friction associated with the gravitational 
wake induced in the medium; and the direct gravitational 
force arising from the disk potential. In typical AGN 
disks, the first two effects dominate for $r \lesssim 10^3\, r_g$, 
while the contribution from the disk potential 
is generally subdominant \cite{Grishin:2023riv}.

For inclined orbits, the secondary crosses the disk 
twice per orbital period. At each crossing, it 
experiences a hydrodynamic drag force
\begin{equation}
\mathbf{F}_{\rm drag} \simeq - \pi\, R_2{}^2\, \rho(r,z)\, v_{\rm rel}\, \mathbf{v}_{\rm rel}\,,
\end{equation}
where $\pi R_2{}^2$ is the cross-sectional area of 
the secondary, and a Gaussian vertical density profile is assumed, namely
\begin{equation}
\rho(r,z) = \frac{\Sigma(r)}{\sqrt{2\pi}\, H} 
\exp\left[-\frac{z^2}{2 H^2}\right]\,.
\end{equation}
The duration of each crossing can be estimated as
\begin{equation}
\Delta t_{\rm cross} \simeq \frac{2 H}{v_{\rm orb}\sin \iota}\,,
\end{equation}
leading to an impulse in the orbital momentum
\begin{equation}
\Delta \mathbf{p}_{\rm drag} \simeq 
- \frac{2 \pi\, R_2{}^2\, \Sigma(r)}{\sin \iota} 
\,\mathbf{v}_{\rm rel}\,.
\end{equation}

At this point, it is necessary to estimate the 
relative importance of the different effects 
induced by the disk, with respect to the central 
gravitational force $a_{\rm grav}=GM/r^2$. For an 
effective drag acceleration over a disk crossing, 
it holds \cite{DeLaurentiis:2022qjq}
\begin{equation}
\frac{a_{\rm drag}}{a_{\rm grav}} \sim 
\frac{3 \kappa_d}{8\pi} \,
\frac{\rho R_2{}^2 v_{\rm rel}^2}{m}
\frac{r^2}{GM}\,,
\end{equation}
where $\kappa_d\sim\mathcal{O}(1)$ is the drag 
coefficient. For instance, stellar-mass secondaries 
embedded in AGN disks satisfy $a_{\rm drag}/a_{\rm grav}\sim10^{-8}-10^{-5}$.

The gravitational wake generated in the gaseous 
medium also gives rise to dynamical friction. 
Following Ostriker's treatment of a collisional 
gaseous medium \cite{Ostriker:1998fa}, the 
corresponding force acting on the secondary can 
be written as
\begin{equation}\label{DF-force}
\mathbf{F}_{\rm DF} =
-4 \pi \rho(r,z)
\left(\frac{G m}{v_{\rm rel}}\right)^2
\mathcal{I}(\mathcal{M})
\, \hat{\mathbf{v}}_{\rm rel}\,,
\end{equation}
where $\mathcal{M}=v_{\rm rel}/c_s$ is the Mach 
number and
\begin{equation}
\mathcal{I}(\mathcal{M})=
\begin{cases}
\frac{1}{2}\ln\!\left(\frac{1+\mathcal{M}}{1-\mathcal{M}}\right)-\mathcal{M},
& \mathcal{M}<1\,,\\[2mm]
\ln\Lambda+\frac{1}{2}\ln\!\left(1-\frac{1}{\mathcal{M}^2}\right),
& \mathcal{M}>1\,.
\end{cases}
\end{equation}
In AGN disks, the motion is typically supersonic 
($\mathcal{M}\gg1$), yielding $F_{\rm DF}\propto \ln\Lambda/v_{\rm rel}^2$ (see, e.g., \cite{Tagawa:2019osr}).

The corresponding acceleration can be compared 
with the central gravitational attraction, 
yielding \cite{Narayan:1999ay}
\begin{equation}\label{DF-grav-ratio}
\frac{a_{\rm DF}}{a_{\rm grav}}
\sim
4\pi \ln\Lambda\,
\frac{G m \rho r^2}{M v_{\rm rel}^2}\,,
\end{equation}
where $\ln\Lambda$ is the Coulomb logarithm. Since
$a_{\rm DF}\propto m$ whereas $a_{\rm drag}\propto R_2^2$, 
dynamical friction generally dominates over 
hydrodynamic drag for compact objects such as 
BHs.

Lastly, the direct gravitational acceleration 
generated by the disk itself simply leads to
\begin{equation}
\frac{a_{\rm disk}}{a_{\rm grav}}
\sim
\frac{M_{\rm disk}(<r)}{M}
\ll 1\,.
\end{equation}
Since AGN disks are typically much less massive 
than the central SMBH, this contribution is 
usually subdominant compared to both hydrodynamic 
drag and dynamical friction at $r\lesssim10^3\,r_g$ \cite{Grishin:2023riv}. Nonetheless, the 
cumulative action of the dominant dissipative 
effects over $\sim10^4$--$10^6$ orbital periods 
may drive a significant secular evolution of 
the orbital parameters.

In addition, the secondary accretes mass from the 
disk during each crossing, further modifying its 
momentum. Modeling the interaction as an inelastic 
collision, momentum conservation implies
\begin{equation}
(m + \Delta m)\mathbf{v}' = m \mathbf{v} + \Delta m \mathbf{v}_{\rm gas}\,.
\end{equation}
The accreted mass $\Delta m$ is estimated using the 
Bondi--Hoyle--Lyttleton prescription \cite{Edgar:2004mk}:
\begin{equation}
\Delta m =
\frac{8 \pi G^2 H \rho\, m^2}
{v_z\left(v_{\rm rel}^2 + c_s^2\right)^{3/2}}\,,
\end{equation}
where $v_z$ is the vertical component of the 
orbital velocity and $c_s$ is the local sound 
speed (consistently defined in Eq.~(\ref{eq:nu})). 
The corresponding velocity change is then
\begin{equation}
\Delta \mathbf{v}
=
\frac{\Delta m}{m+\Delta m}
\left(\mathbf{v}_{\rm gas}-\mathbf{v}\right)\,.
\end{equation}

The hierarchy among the various contributions provides 
a useful guide for the numerical implementation. In particular, the estimates above suggest that only a 
subset of the disk-induced effects needs to be retained 
in order to capture the dominant orbital evolution.

In view of the above estimates, throughout this work we 
treat mass accretion and dynamical friction as the 
dominant mechanisms governing the interaction between 
the secondary and the disk, while neglecting the
comparatively weaker effects discussed above. We 
further adopt a fixed value $\ln\Lambda=3$ in all 
simulations.

\subsection{Particle motion in Kerr geometry}

The Keplerian treatment discussed so far provides a 
useful description of the local forces acting on the 
secondary. However, the conservative component of 
the orbital dynamics is more accurately described 
by geodesic motion in the spacetime generated by the 
central SMBH. For this reason, throughout this work 
we replace Keplerian trajectories by bound timelike 
geodesics in Kerr spacetime, while retaining the same 
effective prescriptions for the interaction between 
the secondary and the disk.

This approximation should be regarded as a hybrid scheme. 
Between two successive disk crossings, the secondary 
follows an exact Kerr geodesic, whereas the momentum 
exchange with the gaseous medium is computed using 
the Keplerian prescriptions introduced in the previous 
section. Such an approach allows us to incorporate the 
dominant relativistic corrections to the orbital motion 
while keeping the environmental coupling computationally 
tractable. A fully relativistic treatment of the disk--secondary interaction would require modeling the hydrodynamic 
and accretion processes directly in the Kerr geometry, 
which is beyond the scope of the present work and will 
be addressed in future investigations.
Finally, we neglect higher-order post-Keplerian terms such as GW emission because they are insignificant, as we show in the Appendix.

\subsubsection*{Orbital motion between crossings}

We follow the analytical framework developed in 
\cite{Fujita:2009bp}, which provides closed-form 
solutions for bound timelike geodesics in Kerr 
spacetime in terms of elliptic functions, whose 
main ingredients are briefly summarized here.

The starting point is the separability of the Hamilton--Jacobi 
equation in Boyer--Lindquist coordinates $(t,r,\theta,\phi)$, 
which implies conservation of the energy $E$, the axial 
component of the angular momentum $L$, and the Carter 
constant $Q$ \cite{Carter:1968ks}, along each geodesic. 
Introducing the \textit{Mino time} $\lambda$ through
\[
d\lambda = \frac{d\tau}{r^2 + \chi^2\cos^2\theta}\,,
\]
where $\chi$ denotes the Kerr spin parameter, the radial 
and polar equations of motion decouple in the form
\begin{eqnarray}
    \left(\frac{dr}{d\lambda}\right)^2 &=& R(r(\lambda))\,, \nonumber \\ 
    \qquad
    \left(\frac{d\theta}{d\lambda}\right)^2 &=& \Theta(\theta(\lambda))\,,
\end{eqnarray}
with
\begin{align}
R(r)
&=
\left[E(r^2+\chi^2)-\chi L\right]^2
\nonumber \label{R-Kerr}\\
&\qquad\qquad\;
-\Delta
\Bigl[
r^2+(L-\chi E)^2+Q
\Bigr]\,,
\\
\Theta(\theta)
&=
Q-\cos^2\theta
\left[
\chi^2(1-E^2)
+\frac{L^2}{\sin^2\theta}
\right]\,. \label{R-Theta}
\end{align}
and $\Delta = r^2 - 2Mr + \chi^2$. The functions 
$R(r)$ and $\Theta(\theta)$ determine the allowed 
regions of motion, fully characterizing the radial 
and polar dynamics. In particular, when restricting 
the analysis to \textit{bound} orbits, the radial 
potential can be factorized as
\begin{equation}
R(r) = (1-E^2)(r_1-r)(r-r_2)(r-r_3)(r-r_4)\,,
\end{equation}
with $r_1 \ge r_2 \ge r_3 \ge r_4$. The radial 
motion can then be expressed analytically as
\begin{equation}
r(\lambda) =
\frac{
r_3(r_1-r_2)\,\mathrm{sn}^2(u_r(\lambda),k_r)
-r_2(r_1-r_3)
}{
(r_1-r_2)\,\mathrm{sn}^2(u_r(\lambda),k_r)
-(r_1-r_3)
}\,,
\end{equation}
where $\mathrm{sn}(u,k)$ denotes the Jacobi elliptic 
function and
\begin{eqnarray}
u_r(\lambda) &=& 
\frac{\sqrt{(1-E^2)(r_1-r_3)(r_2-r_4)}}{2}\,\lambda\,, \\
k_r &=& 
\sqrt{\frac{(r_1-r_2)(r_3-r_4)}
{(r_1-r_3)(r_2-r_4)}}\,.
\end{eqnarray}

A similar construction applies to the polar motion. 
In fact, by defining $z=\cos^2\theta$, the polar 
potential can be parameterized in terms of two 
turning points, $z_+$ and $z_-$. The solution of 
the polar equation yields
\begin{equation}
\cos\theta(\lambda) =
\sqrt{z_-}\,\mathrm{sn}(u_\theta,k_\theta)\,,
\end{equation}
with
\begin{equation}
u_\theta =
\sqrt{\chi^2(1-E^2)z_+}\,\lambda\,,
\qquad
k_\theta =
\sqrt{\frac{z_-}{z_+}}\,.
\end{equation}
This representation is particularly useful, as it makes 
explicit the periodic nature of the polar motion in 
Mino time. Once the radial and polar trajectories are 
known, the azimuthal and temporal components follow 
from the first-order equations
\begin{align} \label{eq_sep_Kerr}
\frac{d\phi}{d\lambda} &= 
\Phi_r(r(\lambda)) + \Phi_\theta(\theta(\lambda))\,,\nonumber \\
\frac{dt}{d\lambda} &= 
T_r(r(\lambda)) + T_\theta(\theta(\lambda))\,,
\end{align}
where $\Phi_r$ and $T_r$ are functions only of $r$, 
while $\Phi_\theta$ and $T_\theta$ depend exclusively 
on $\theta$. Explicit expressions for these functions 
follow directly from the separated Kerr geodesic 
equations \cite{Fujita:2009bp} (see also \cite{Schmidt:2002qk}). 
The solutions of Eqs.~(\ref{eq_sep_Kerr}) can then be written 
schematically as\footnote{Each contribution can be evaluated analytically and expressed in terms of elliptic integrals of 
the first, second, and third kinds.}
\begin{equation}
\phi(\lambda) = \phi_r(\lambda) + \phi_\theta(\lambda)\,,
\qquad
t(\lambda) = t_r(\lambda) + t_\theta(\lambda)\,.
\end{equation}
%

\section{Numerical implementation}
\label{Sec-Numerical-Implementation}

In this section, we describe the numerical setup 
employed throughout the simulations. In particular, 
we discuss the initialization procedure, the 
orbital updates applied after each interaction 
with the disk, and the set of quantities used to 
characterize the orbital evolution.

\subsection{Initialization and evolution updates}

We initialize each orbit by specifying the triplet
$(a_0,e_0,x_0)$, corresponding to the initial values 
of the semi-major axis, eccentricity, and orbital 
inclination $\iota_0$ with respect to the equatorial 
plane, through $x_0=\cos\iota_0$. This parametrization 
is in \textit{one-to-one} correspondence with the set $(E,L,Q)$, 
together with the normalization condition for the 
four-velocity. The explicit relation between both sets of parameters is detailed in Appendix \ref{app:keplerian-kerr-conversion}.

In addition, we fix the initial
argument of pericentre to $\omega_0=-\pi/3$ in all our
simulations. We adopt the initial true anomaly
$\vartheta_0=\pi/3$, such that
$\omega_0+\vartheta_0=0$. Thus, the initial position of
the secondary lies at the intersection of the orbital and
equatorial planes, corresponding to a disk crossing.
This choice follows the convention adopted in
Ref.~\cite{Spieksma:2025wex}, allowing for a direct
comparison with the results presented there. In general,
disk crossings occur at true anomalies $\vartheta_{\rm cross}=-\omega$ and $\vartheta_{\rm cross}=\pi-\omega$, and the corresponding crossing radii are
\begin{equation}
    r_{\pm}
    =
    \frac{a(1-e^2)}
    {1\pm e\cos\omega}.
\end{equation}
For our choice of $\omega_0=-\pi/3$, this gives
\begin{equation}
    r_{\pm}
    =
    \frac{a(1-e^2)}
    {1\pm e/2}.
\end{equation}
The two disk crossings therefore occur at distinct radii,
with their separation increasing with eccentricity.
Consequently, the secondary encounters the disk at
different local densities, sound speeds, and relative
velocities at the two crossings, resulting in different
accretion and dynamical-friction kicks during each orbit\footnote{
We do not explore the dependence of the orbital evolution
on $\omega$ in the present work, and leave such a systematic
study for future investigation.}.

We initialize the orbital velocity using the standard 
Keplerian relations,
\begin{align}
    v_r &= \sqrt{\frac{M}{a(1-e^2)}}\,e \sin\vartheta,\\
    v_\theta &= \sqrt{\frac{M}{a(1-e^2)}}\,(1 + e \cos\vartheta),
\end{align}
while $v_\phi = 0$ for a planar orbit. These expressions 
follow from $v_r=\dot r$ and $v_\theta=r\dot\vartheta$, 
with $r=\lvert\mathbf{r}(t)\rvert$ given by 
Eq.~(\ref{rposition-keper}). In a purely Keplerian treatment, 
such as that developed in \cite{Spieksma:2025wex}, once the 
orbital parameters after a disk crossing are known, the 
location of the next crossing can be determined analytically 
from the intersection of the Keplerian orbit with the disk plane.

In the present work, however, the orbital motion between 
successive interactions is described by Kerr geodesics. 
As a consequence, the trajectory is specified through the 
functions $r(\lambda)$ and $\theta(\lambda)$ rather than 
by a fixed Keplerian ellipse, making the location of the 
next equatorial crossing a numerical problem. We therefore 
determine the next crossing by searching for the roots of
\begin{equation}
\theta(\lambda)-\frac{\pi}{2}=0\,,
\end{equation}
corresponding to intersections of the orbit with the equatorial 
plane. Our implementation samples $\theta(\lambda)$ on a 
numerical grid, identifies sign changes of 
$\theta(\lambda)-\pi/2$, and refines the root using a Brent 
root-finding algorithm.

\begin{figure*}
  \centering
  \includegraphics[scale=0.38]{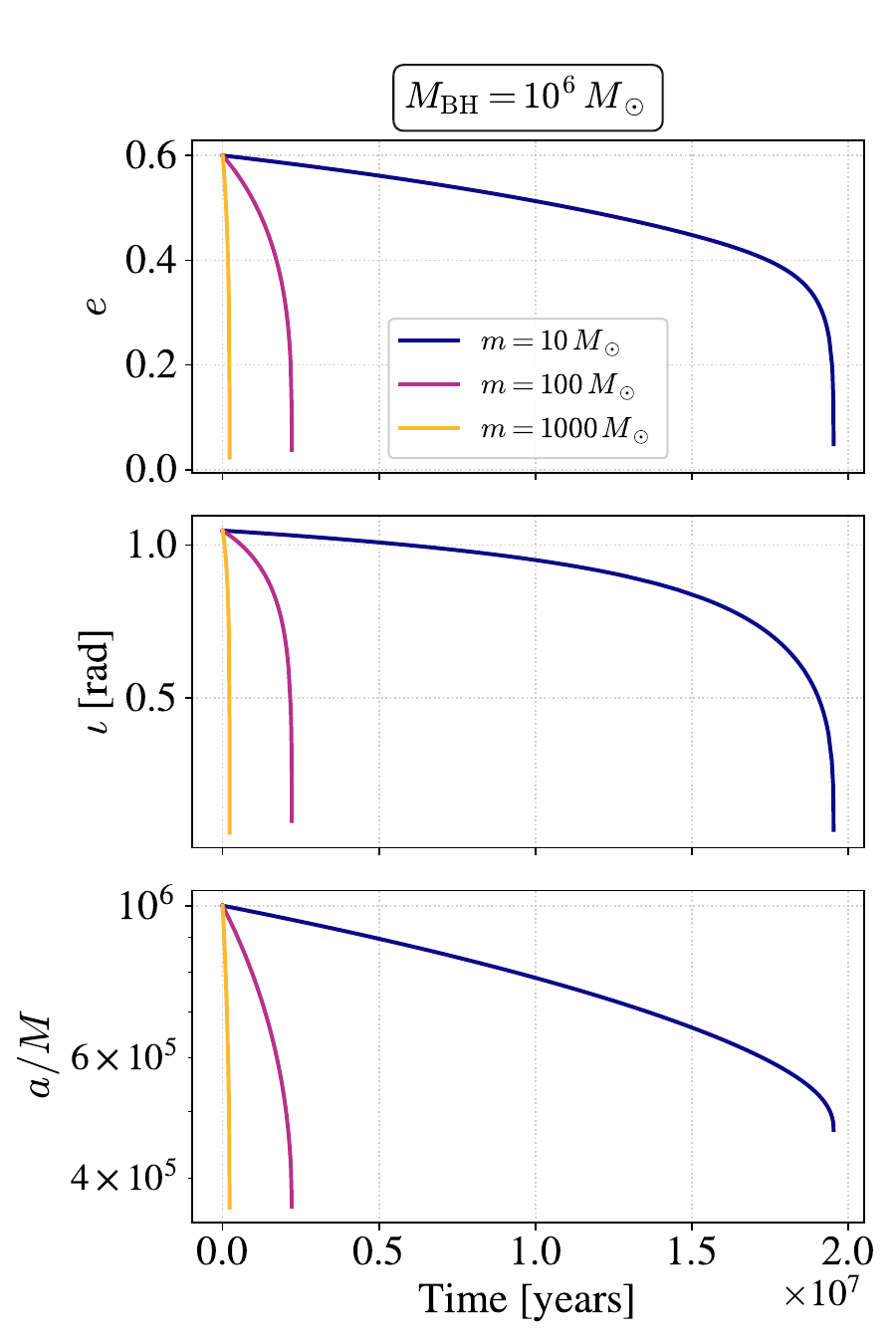}
  \includegraphics[scale=0.38]{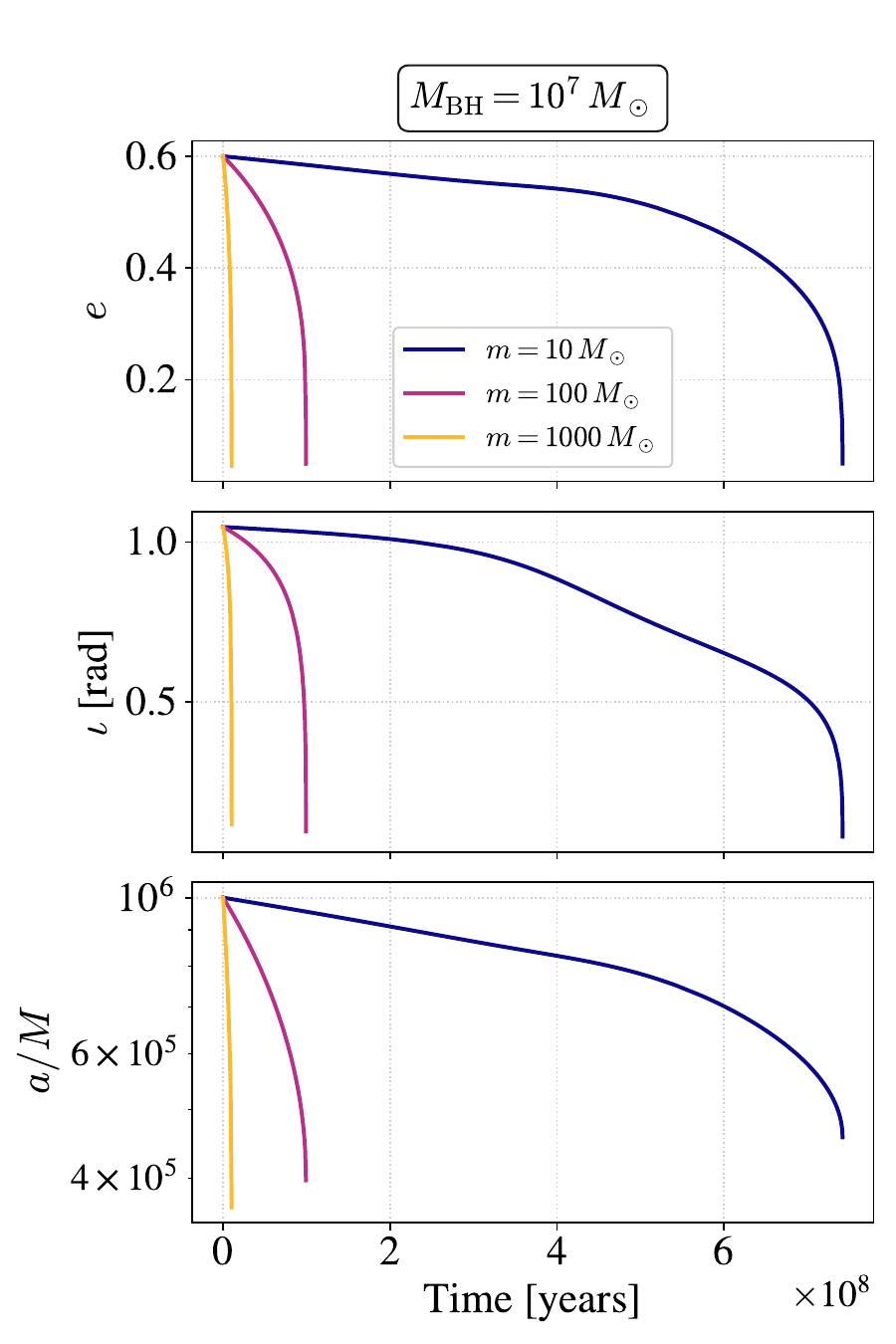}
  \includegraphics[scale=0.38]{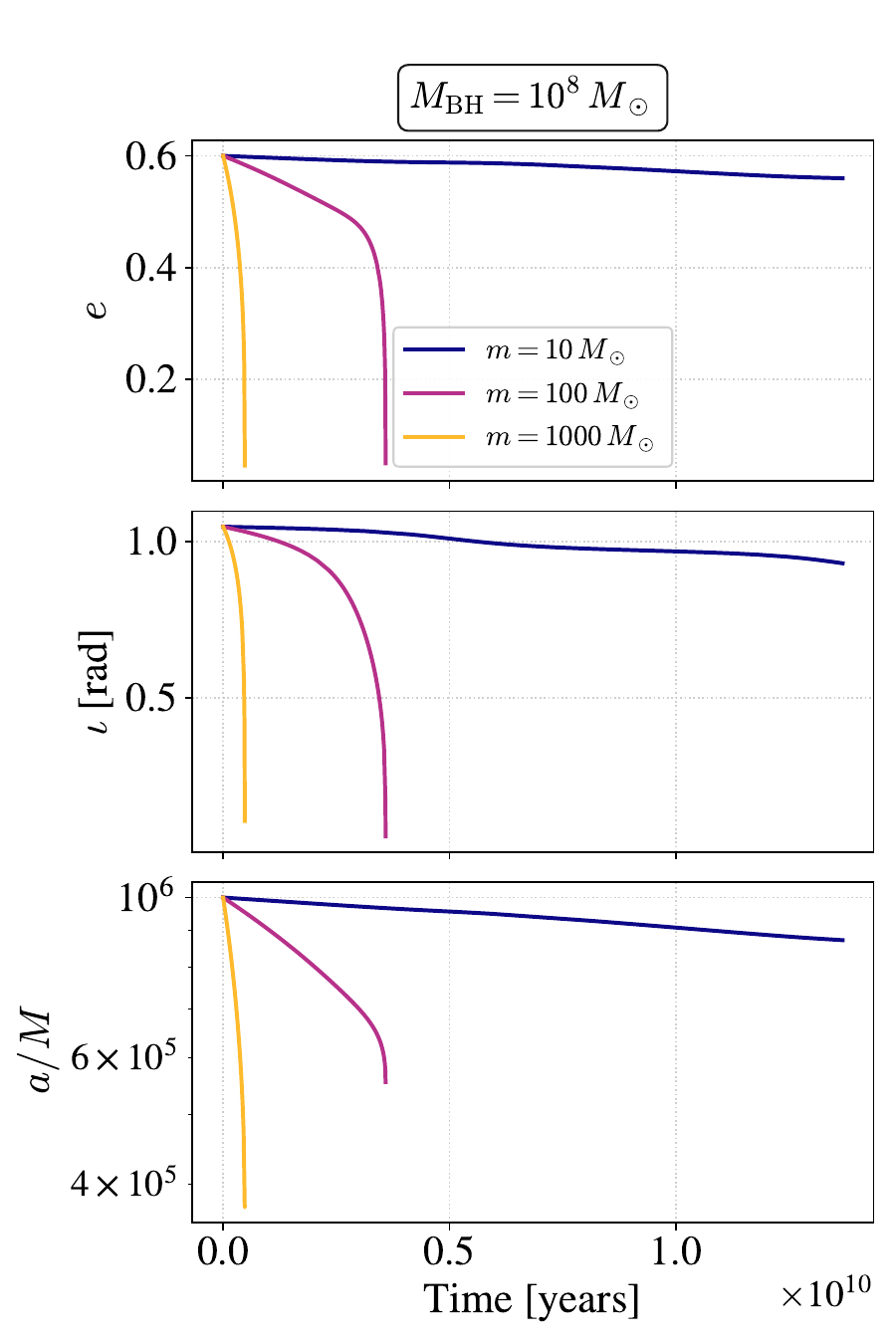}
  \caption{
\textit{Orbital evolution of the secondary.}
Time evolution of the orbital parameters (eccentricity $e$, 
orbital inclination $\iota$, and normalized semi-major 
axis $a/M$), for a set of nine configurations. Each column 
(from left to right) corresponds to different central BH 
masses, $M_{\rm BH} = 10^6,\,10^7,\,10^8\,M_\odot$ 
respectively, while colors distinguish the secondary masses 
$m = 10,\,10^2,\,10^3\,M_\odot$. All systems are initialized 
at a separation $a_0 = 10^6\,M$ from the central BH, and 
evolved in a Schwarzschild spacetime including disk-induced
accretion and dynamical-friction effects.
The dynamics exhibit a two-stage behavior: a rapid decrease 
of the orbital inclination, driving the system toward a 
nearly coplanar configuration with the disk, followed by a 
slower circularization driven by dissipative interactions 
during repeated disk crossings. The characteristic 
circularization timescale is $\tau \sim 10^7$ yr, on average.
}
\label{Fig2-dynamics-evo}
\end{figure*}

For the disk interactions, we follow the framework described 
in the previous section. The density $\rho(r)$, sound speed 
$c_s(r)$, and scale height $H(r)$ are interpolated from the 
selected disk model and evaluated at each equatorial crossing. 
Similarly to the treatment adopted in \cite{Spieksma:2025wex}, 
the velocity is modified by dynamical friction according to
\begin{equation}
    \Delta \vec{\mathbf{v}}
    =
    \frac{\vec{\mathbf{F}}_{\rm DF}\,\Delta t}{m}\,,
\end{equation}
where $\vec{\mathbf{F}}_{\rm DF}$ is given by 
Eq.~(\ref{DF-force}), and the interaction time is estimated as
\begin{equation}
    \Delta t=\frac{2H}{v_\theta}\,.
\end{equation}

At each disk crossing, the orbital parameters are recorded. 
For the large semi-major axes considered in our simulations, 
the orbital inclination evolves slowly, and the parameter 
$x=\cos\iota$ remains a useful proxy for tracking the 
orientation of the orbit. Following the interaction with 
the disk, we reconstruct the four-velocity of the secondary 
and compute the corresponding constants of motion $(E,L,Q)$, 
or equivalently the updated set of orbital parameters. The 
geodesic is then reinitialized and evolved until the next 
disk-crossing event.

The orbital evolution between successive disk crossings is 
computed using \texttt{KerrGeoPy}, a Python package for the 
accurate calculation of bound timelike geodesics in Kerr 
spacetime \cite{Park:2024sjj}. The package implements the 
analytical formalism based on Mino-time parametrization and 
elliptic integrals, enabling a fast and accurate evaluation 
of generic bound trajectories together with their associated 
fundamental frequencies and spacetime coordinates.

Throughout the evolution, we record the orbital parameters 
and constants of motion after each disk crossing. In 
particular, we monitor the semi-major axis, eccentricity, 
inclination, and the corresponding set of Kerr integrals 
$(E,L,Q)$, together with the mass growth of the secondary.

Finally, the simulation is terminated if the eccentricity 
falls below a prescribed threshold $e_{\min}$ (in our case, 
$e_{\min}=10^{-3}$), indicating a quasi-circular orbit\footnote{By the time $e<e_{\min}$ is reached, the orbit has already
undergone the faster alignment phase and is effectively coplanar with
the disk. The secondary is then embedded in the disk, so discrete
crossings cease to be well defined and quantities such as $\Delta t$
diverge.}; or when the orbit becomes unbound 
or plunging, usually signaled by the failure of the conversion 
from constants of motion to orbital parameters.

\section{Results}
\label{Sec-Results}

\begin{figure*}
  \centering
  \includegraphics[scale=0.5]{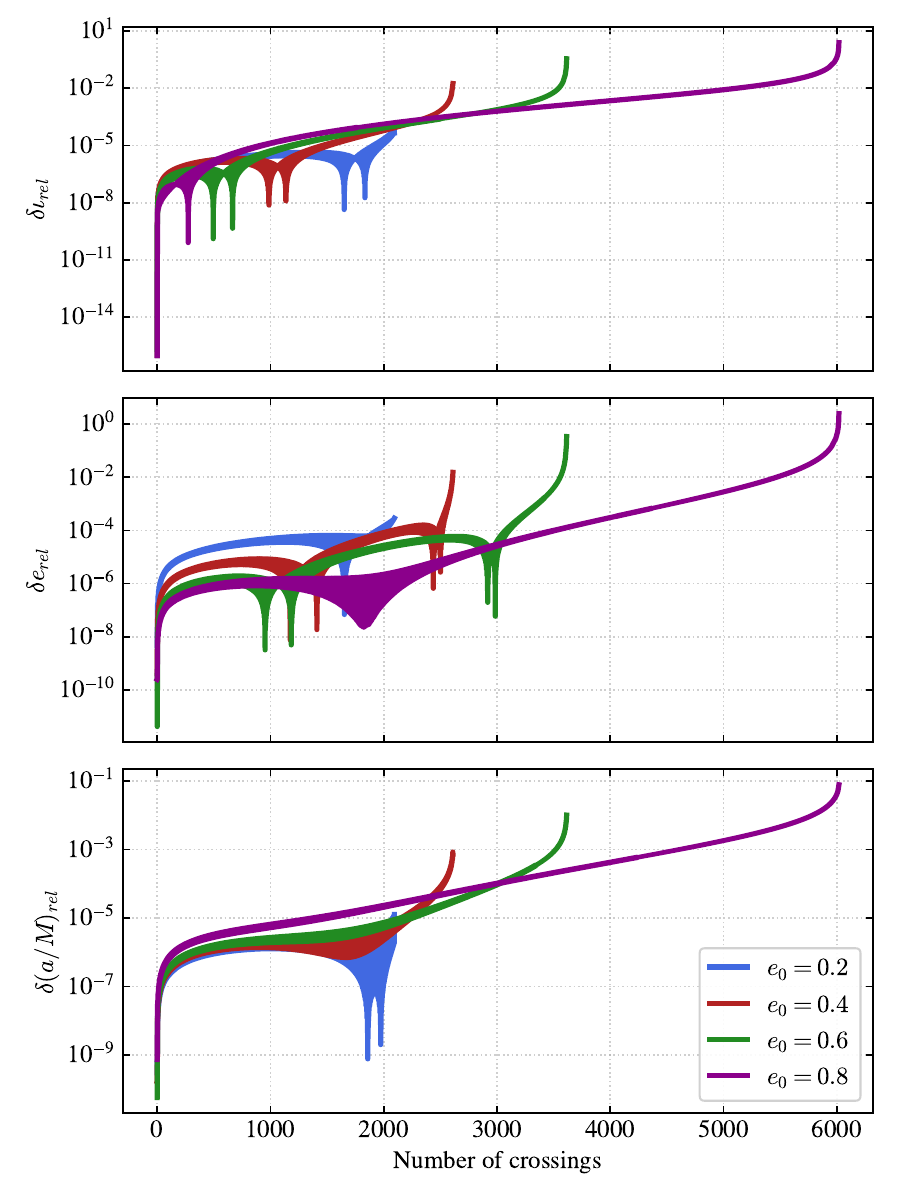}
  \includegraphics[scale=0.5]{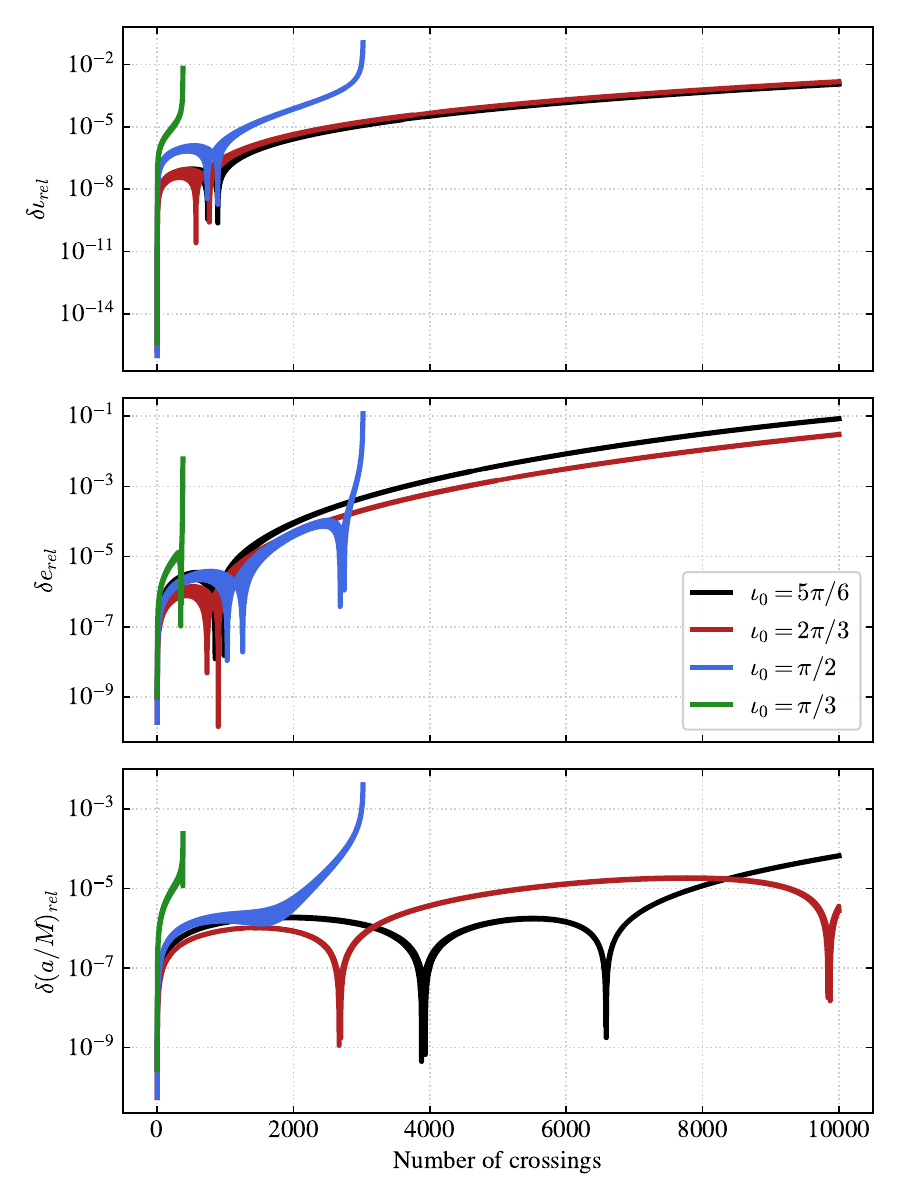}
 \caption{
\textit{Comparison with a Keplerian treatment.}
Relative logarithmic difference between the Keplerian 
and relativistic evolutions of binaries embedded in a 
SG disk. Two sets of initial conditions are considered. 
In the left panel, the initial eccentricity is varied 
over $e_0 \in \{0.2, 0.4, 0.6, 0.8\}$ while keeping 
the inclination fixed at $\iota_0=\pi/3$. In the right 
panel, the initial inclination $\iota_0$ is varied 
while the eccentricity is fixed to $e_0=0.5$. In all 
cases, the initial separation is $a_0 = 10^6 M$, and 
the initial true anomaly is set to $\theta_0 = \pi/2$. 
The relative discrepancy grows throughout the evolution, 
revealing systematic differences between the Keplerian
evolution and the Kerr-based treatment even for 
configurations initialized at large orbital 
separations.}
\label{Fig3-error-comparison-Keplerian}
\end{figure*}

We now present the main results of our work. We begin by investigating the orbital evolution of the secondary at large distance from the primary BH, highlighting the differences between relativistic and Keplerian treatments. We then extend the study to the intermediate-separation regime, where relativistic effects become important. We also examine the dependence on 
the disk model, as well as with the spin of the central 
BH, assessing their impact on the efficiency of orbital 
circularization and alignment.

\subsection{Evolution at large separation}
\label{Sec-ResA}

We start by studying the evolution of the secondary 
orbit around a non-rotating Schwarzschild BH, 
assuming a SG disk model. This setup facilitates a 
direct comparison with previous Keplerian studies 
\cite{Spieksma:2025wex}. We consider different values of the primary 
and secondary masses, namely
$M_{\rm BH} = \{10^6,\,10^7,\,10^8\}\, M_{\odot}$ and
$m = \{10,\,10^2,\,10^3\}\, M_{\odot}$, respectively.

We follow the evolution of the eccentricity $e$, 
inclination $\iota$, and semi-major axis $a/M$ of 
the secondary orbit throughout the evolution, 
starting from an initial separation $a_0 = 10^6\,M$. For all configurations considered, the orbit exhibits 
a clear tendency towards circularization. Local 
interactions with the disk, together with the 
exchange of mass and linear momentum between the 
secondary and the surrounding gas, drive a gradual 
dissipation of the initial eccentricity.
For the range of primary and secondary masses displayed 
in Fig.~\ref{Fig2-dynamics-evo}, we report a characteristic 
circularization timescale of order $\tau \sim 10^7$ 
years. We also find that increasing the mass of the 
secondary generally leads to longer circularization 
timescales.

\begin{figure*}
  \centering
  \includegraphics[scale=0.61]{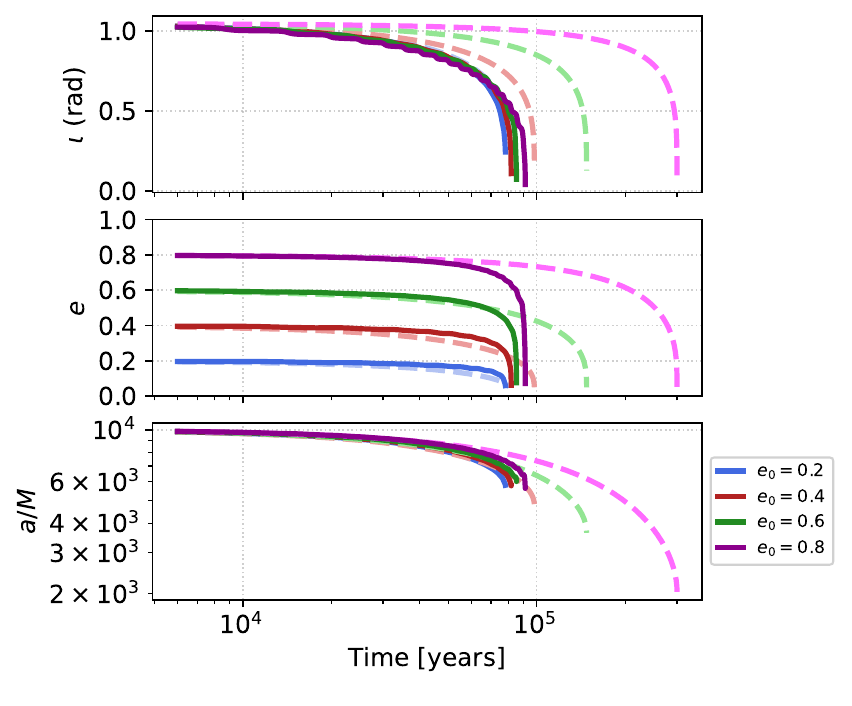}
  \includegraphics[scale=0.61]{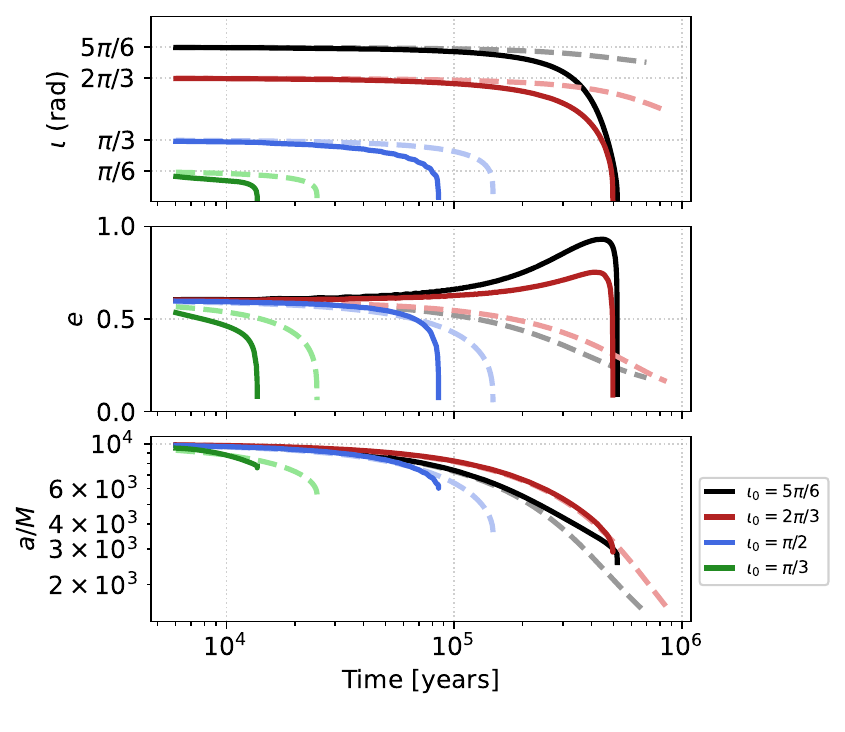}
\caption{\textit{Dynamics in the intermediate region.}
Time evolution of the inclination $\iota$ (top), eccentricity $e$ (middle), and semi-major axis $a$ (bottom) of the secondary 
orbit due to its interaction with the accretion disk. The mass of the secondary is set to $m=10\, \Msun$, while the initial separation is $a_0 = 10^4 M$. Solid curves 
correspond to evolutions computed on a Kerr background, 
while dashed curves denote the corresponding Keplerian 
treatment. The left column shows sequences with fixed 
initial inclination $\iota_0=\pi/3$ and varying eccentricities 
$e_0 \in \{0.2,0.4,0.6,0.8\}$, whereas the right column 
displays configurations with fixed eccentricity $e_0=0.6$ 
and initial inclinations $\iota_0 \in \{\pi/3,\pi/2,2\pi/3,5\pi/6\}$. Noticeable deviations from the Keplerian evolution emerge 
already at early times, revealing systematic differences 
in the dissipation rates and secular evolution of the orbit. 
The effect becomes increasingly pronounced for highly 
eccentric and highly inclined configurations, where the 
geometry and frequency of the disk crossings are more 
strongly affected by relativistic corrections.}
  \label{Fig4-intermediate-dynamics}
\end{figure*}

In addition to the eccentricity damping, the orbital plane gradually aligns with the accretion disk. Interestingly, the alignment timescale is typically about \textit{an order of magnitude shorter} than the circularization timescale, as can also be seen in Fig.~\ref{Fig2-dynamics-evo}. Here, the quoted alignment timescale refers specifically to the \textit{initial fast-relaxation stage}, during which the inclination decreases rapidly while the eccentricity remains of the same order as its initial value. This naturally leads to a two-stage evolution. The orbit first evolves towards a nearly coplanar configuration while retaining a significant eccentricity, although exact alignment is generally not achieved during this initial stage. Only at later times, once the orbit is nearly coplanar with the disk, it undergoes a \textit{more gradual} circularization driven by dissipative interactions with the surrounding gas. For the present initial conditions, the ratio $e_0/e_{\min}\sim 10^2$--$10^3$ implies that the subsequent slow evolution drives the inclination to values \textit{far below} those reached during the initial relaxation, rendering the orbit \textit{effectively coplanar} by the time the circularization threshold $e_{\min}=10^{-3}$ is reached. This also motivates our choice of $e<e_{\min}$ as the termination criterion, as although the inclination does not reach exactly zero in finite time, by the time this eccentricity threshold is reached the secondary is \textit{already effectively embedded} within the disk, and the assumption of discrete, localized disk crossings ceases to be physically meaningful. A geometrical argument underlying this behavior is discussed in Appendix~\ref{Sec-AppendixLast}, where we show that the orbital shrinkage accompanying circularization naturally leads to the embedded regime near the termination of the evolution.

The configurations shown in Fig. (\ref{Fig2-dynamics-evo}) correspond to relatively \textit{large} 
orbital separations, where one might expect a Keplerian 
description to provide an adequate approximation 
to the BH--disk--secondary dynamics. Indeed, the 
gravitational field is weak enough that relativistic 
corrections are often assumed to be negligible. Nevertheless, 
our results show that even mild relativistic effects 
associated with the spacetime geometry of the central 
BH accumulate over the long evolution timescales 
considered here, producing systematic deviations with 
respect to the Keplerian description.

This behavior is illustrated in Fig.~\ref{Fig3-error-comparison-Keplerian}, 
which shows the relative difference between the Keplerian and relativistic evolutions. 
For this analysis, we consider two sets of initial conditions. 
In the first one, the eccentricity is varied over the range 
$e_0 \in \{0.2,0.4,0.6,0.8\}$ while keeping the inclination 
fixed at $\iota_0=\pi/3$ (left panel). In the second, the initial inclination $\iota_0$ is varied 
while the eccentricity is fixed to $e_0=0.5$ (right panel). 
All simulations are initialized with an orbital 
separation $a_0=10^6\,M$ and true anomaly $\theta_0=\pi/2$. 
These configurations encompass the region of parameter 
space previously explored in \cite{Spieksma:2025wex} within a purely 
Keplerian framework.

We find that the relative differences do not remain 
bounded during the evolution, but instead grow 
systematically as the number of crossings between 
the secondary and the accretion disk increases. This 
behavior suggests that even small relativistic 
corrections can accumulate over long timescales, 
producing progressively larger deviations in both 
the orbital geometry and the dissipative processes 
associated with the environment. Consequently, the accuracy of a purely Keplerian 
description appears to deteriorate well before the system 
enters a strongly relativistic regime.

Furthermore, both the magnitude of the discrepancy and 
its growth rate depend sensitively on the initial orbital 
configuration. In particular, systems with larger 
eccentricities exhibit more pronounced deviations, 
since the secondary experiences stronger variations 
in both orbital velocity and radial distance during 
each revolution. This enhances the relative impact of 
relativistic corrections during passages closer to 
the central BH.

A similar trend is observed when varying the initial 
inclination. Changes in $\iota_0$ modify both the 
frequency and geometry of the disk crossings, thereby 
altering the cumulative mismatch between the Keplerian 
and relativistic descriptions. Taken together, these 
results indicate that, even for initially distant 
configurations with $a_0=10^6\,M$, a purely Keplerian 
treatment may be insufficient to accurately capture 
the secular evolution of systems undergoing repeated 
disk--object interactions over long integration times.

\subsection{Evolution at intermediate separations}
\label{Sec-ResB}

In light of the discrepancies identified at large orbital 
separations, we now move to configurations that probe a 
more relativistic region of the orbital parameter space. 
Specifically, 
we consider systems initialized at $a_0 = 10^4 M$, which 
we refer to as the \textit{intermediate regime}. At these 
separations, the interaction with the disk remains 
significant, while relativistic effects associated with 
the spacetime geometry of the central BH exert a 
more noticeable influence on the orbital evolution.

Figure~\ref{Fig4-intermediate-dynamics} 
shows the time evolution of the inclination (top panel), 
eccentricity (middle panel), and semi-major axis (bottom 
panel), comparing trajectories obtained from a Kerr-geodesic
evolution (solid lines) with those derived from Kepler's laws (dashed lines). Two sets of initial conditions are considered once again. 
The left panel explores variations in the initial 
eccentricity ($e_0 \in \{0.2,0.4,0.6,0.8\}$) while keeping 
the inclination fixed at $\iota_0=\pi/3$, whereas the right 
panel examines different initial inclinations, $\iota_0 \in \{\pi/3,\pi/2,2\pi/3,5\pi/6\}$, for a fixed eccentricity $e_0=0.6$.

This regime is particularly interesting because it
combines repeated interactions with the accretion disk within a stronger gravitational field. As a result, geometric effects arising from the curved spacetime around the central BH begin to noticeably influence the orbital dynamics, as already suggested by the trajectories shown in Fig.~\ref{Fig3-error-comparison-Keplerian}.

In contrast to the large-separation case, the discrepancies 
between the relativistic and Keplerian evolutions emerge 
considerably earlier in the intermediate regime and grow more 
rapidly throughout the evolution. The orbital variables 
exhibit systematic differences in both the dissipation 
timescales and the geometry of the trajectories, indicating 
that relativistic effects play a progressively more important 
role in shaping the dynamics of the system.

In particular, the most eccentric configurations display a 
faster evolution and a more pronounced divergence between 
the two descriptions. This behavior can be attributed to 
the repeated close passages near the central BH, 
which enhance the impact of relativistic corrections on the 
exchange of energy and angular momentum. Similarly, variations 
in the initial inclination modify both the frequency and 
geometry of the disk crossings, leading to appreciable 
differences in the orbital damping rate and in the 
inclination evolution.

Overall, these results suggest that the intermediate regime 
marks the onset of a transition in which relativistic 
effects become increasingly relevant for the long-term 
evolution of the system. Consequently, relativistic corrections 
must be properly accounted for well before the secondary 
reaches the innermost regions of the BH gravitational 
potential.

\subsection{Comparison between disk prescriptions}

\begin{figure}
  \centering
  \includegraphics[scale=0.65]{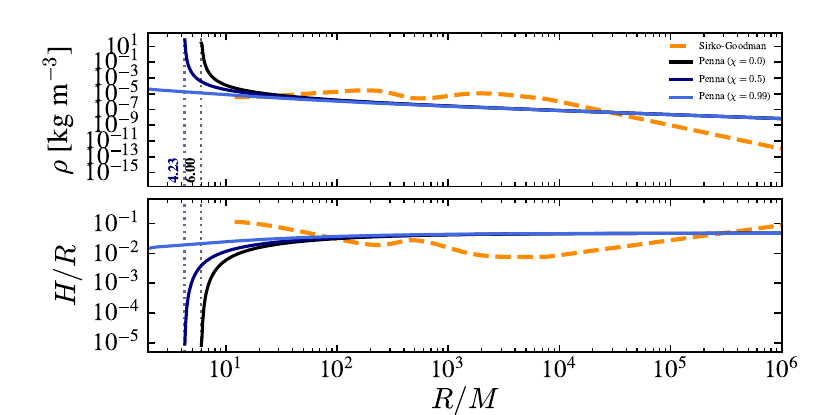}
  \includegraphics[scale=0.65]{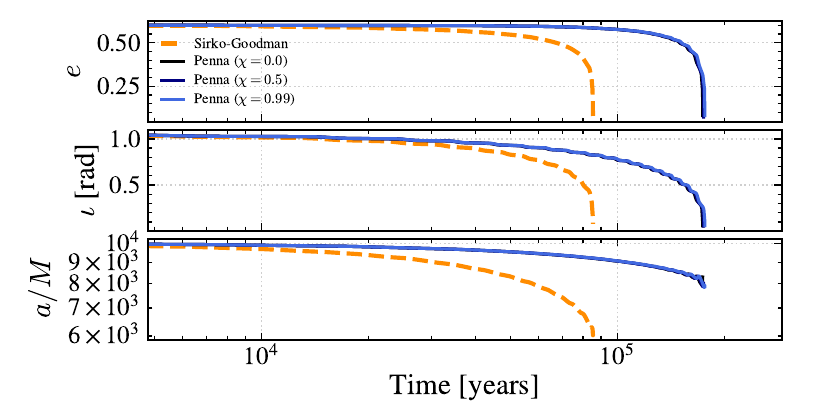}
\caption{\textit{Impact of disk models on the orbital dynamics.} 
Comparison between the SG and Penna disk prescriptions 
and their impact on the evolution of an embedded secondary. 
\textit{Top panel:} midplane density $\rho$ and aspect 
ratio $H/R$ as functions of $R/M$. Dashed curves correspond 
to the SG model, while solid blue curves show the Penna 
solution for different values of the BH spin parameter 
$\chi$. The Penna model predicts lower densities and a 
thicker disk in the region $10^2M \lesssim R \lesssim 10^4M$,
whereas spin-dependent effects become appreciable only
for $R \lesssim 10^2M$. \textit{Bottom panel:} 
evolution of the orbital parameters for a secondary 
initialized at $a_0=10^4M$ with eccentricity $e_0=0.6$. 
The Penna disk (solid blue) produces a slower orbital 
decay and eccentricity damping than the SG model (dashed 
orange), consistent with its lower densities and larger 
scale heights.}
\label{Fig5-SGvsPenna}
\end{figure}

Having established the impact of relativistic corrections 
on the orbital dynamics, we now investigate the dependence 
of the evolution on the adopted accretion-disk model. To 
this end, we compare simulations performed using the SG 
and NT/Penna disk prescriptions introduced in Sec.~\ref{Sec-Preliminaries}. 
Since both models are initialized using the 
same global astrophysical parameters, any difference 
between the resulting trajectories can be attributed 
directly to differences in the underlying disk structure. 
In particular, the distinct radial profiles of the 
density, sound speed, and scale height modify both the 
strength of the disk--secondary interaction and the 
cumulative exchange of mass and momentum throughout 
the evolution.

The top panel of Fig.~\ref{Fig5-SGvsPenna} compares the radial 
structure of the two disk prescriptions through the midplane 
density $\rho$ and geometric aspect ratio $H/R$ as functions 
of the dimensionless radius $R/M$. The dashed curves 
correspond to the SG model, whereas the solid 
blue curves represent the Penna solution computed for 
different values of the BH spin parameter.

The two prescriptions exhibit systematic variations throughout 
the radial domain of interest. In particular, the Penna model 
predicts lower midplane densities in the 
intermediate region $10^2M \lesssim R \lesssim 10^4M$, 
together with a geometrically thicker disk, as reflected by 
the larger values of $H/R$. These differences arise from the 
relativistic corrections incorporated in the disk structure, 
which modify the radial distribution of pressure, temperature, 
and viscous dissipation with respect to the purely Keplerian SG model.

The influence of the BH spin becomes appreciable only 
at radii $R \lesssim 10^2M$, where strong-field effects 
associated with the Kerr geometry begin to alter the disk 
properties. At larger distances, the various Penna solutions progressively converge toward a common behavior, although 
differences with respect to the SG model persist in both 
the density and thickness profiles.

These structural differences are particularly relevant for 
the evolution of the secondary, since the efficiency of 
disk--object interactions depends sensitively on both the 
local gas density and the vertical structure of the 
disk. Consequently, even modest changes in the underlying 
disk prescription can lead to appreciable differences in 
the secular evolution of the orbit.

The bottom panel of Fig.~\ref{Fig5-SGvsPenna} illustrates 
the dynamical consequences of these structural differences 
on the evolution of a secondary initially located at 
$a_0=10^4M$ with eccentricity $e_0=0.6$. We show the 
time evolution of the main orbital parameters under the action of the dissipative effects generated by each disk model. The dashed orange curves 
correspond to the SG prescription, 
while the solid blue curves represent the evolution 
obtained using the relativistic NT/Penna disk structure.

The two models lead to quantitatively different evolutionary 
tracks. The SG prescription predicts a more efficient 
extraction of orbital energy and angular momentum, 
resulting in a faster circularization of the orbit and a 
more rapid decrease of the semi-major axis. By contrast, the 
Penna model produces a slower evolution of all orbital 
parameters.

\begin{figure}
  \centering
  \includegraphics[scale=0.4]{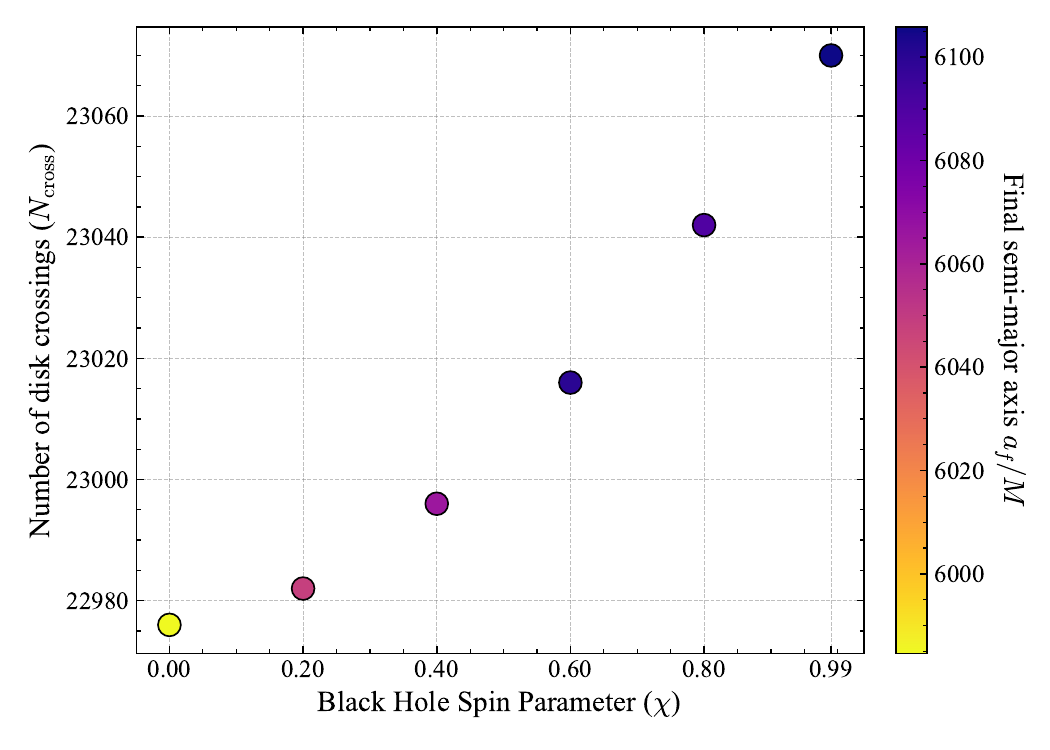}
  \includegraphics[scale=0.4]{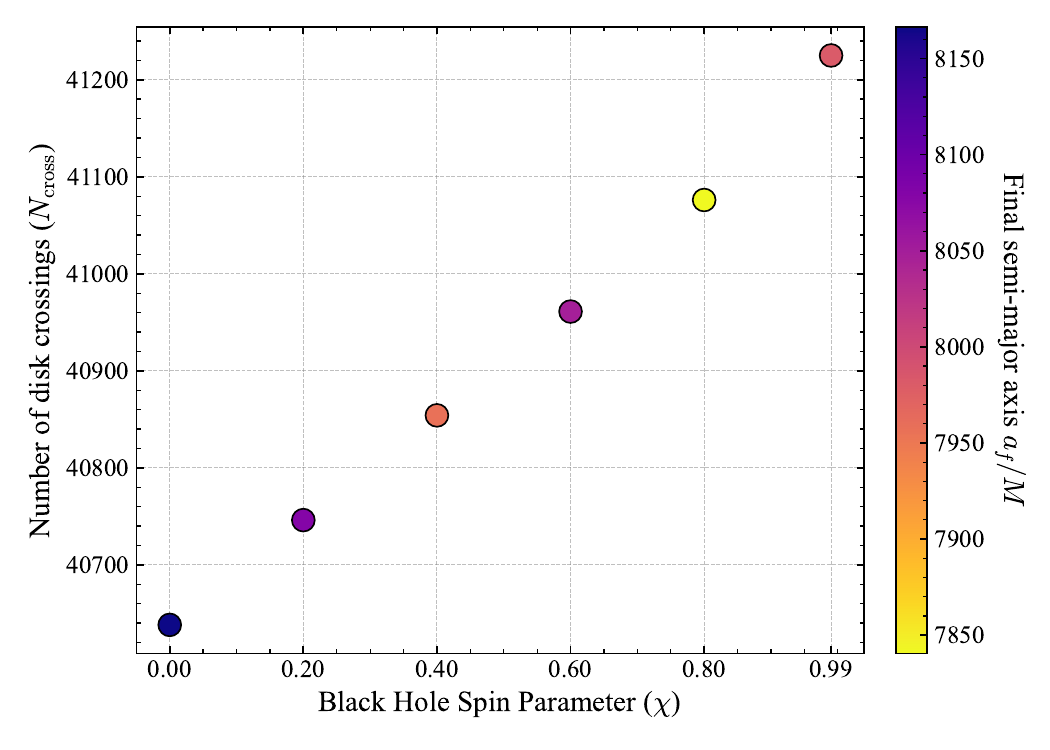}
\caption{\textit{Spin influence on the circularization 
efficiency.} Number of disk crossings 
$N_{\rm cross}$ required for the eccentricity to 
decrease below the threshold $e_f<0.05$, starting 
from identical initial orbital configurations, 
for the SG (top) and Penna (Novikov--Thorne) (bottom) 
disk models. The color scale indicates the final 
semi-major axis, in units of the central BH 
mass, of the circularized orbit. In both disk 
prescriptions, the circularization efficiency 
exhibits only a weak dependence on the spin parameter 
$\chi$, with variations in $N_{\rm cross}$ remaining 
below the percent level. 
This indicates that the circularization efficiency 
depends more strongly on the adopted disk structure 
than on the spin of the central BH.}.
\label{Fig6-spin-dependence}
\end{figure}

This difference can be traced back to the lower densities 
and larger scale heights predicted by the relativistic 
disk model. Since the efficiency of the disk--secondary 
interaction depends sensitively on the local gas density and 
on the vertical structure of the medium crossed by the 
compact object, even moderate differences in the underlying 
disk properties accumulate over long timescales and produce 
appreciable changes in the orbital evolution.

Our results indicate that the adopted disk prescription 
can have a significant impact on the predicted evolution 
of compact objects embedded in AGN disks, particularly 
when repeated disk crossings are involved.

Finally, we investigate the dependence of the 
circularization efficiency on the spin parameter $\chi$ 
of the central BH for both disk prescriptions 
considered in this work. For each configuration, we 
compute the number of disk crossings, $N_{\rm cross}$, 
required to reduce the orbital eccentricity below 
the threshold $e_f<0.05$, thereby reaching a quasi-circular 
state. The results are reported in Fig.~\ref{Fig6-spin-dependence}, 
where the color scale indicates the final semi-major axis, 
in units of the central BH mass, associated with 
the circularized orbit.

For the SG model (top panel), the number of crossings 
required to achieve circularization is found to be 
remarkably insensitive to the value of $\chi$. Although 
the physical timescale associated with the evolution 
depends on the spin through the orbital frequencies 
and disk properties, the total dispersion in 
$N_{\rm cross}$ remains below $0.3\%$. This indicates 
that the overall efficiency of the circularization 
process is largely unaffected by the rotation of the 
central BH.

A similar behavior is observed for the Penna model (bottom 
panel). Despite the fact that the disk structure itself 
depends on the Kerr geometry, the number of crossings 
required to reach the circularized state varies only 
weakly across the range of spins considered. The impact 
of the spin therefore remains modest even when 
relativistic corrections significantly affect the 
properties of the accretion flow.

Taken together, these results indicate that the structure
of the accretion disk plays a substantially larger role in
determining the circularization efficiency than the spin of
the central BH. The process is primarily controlled
by the cumulative interaction between the secondary and the
surrounding gaseous medium, while spin-dependent relativistic
effects provide only a secondary correction.

\subsection{Comparison with previous studies}

We now compare the results here presented with recent studies of EMRIs interacting with accretion disks; particularly with Refs. \cite{fxnd-r4bs} and \cite{10.1093/mnras/staa3004}. We find this comparison very useful as it provides a consistency check on our results and, at the same time, allows us to identify the effects that arise from extending the standard disk-crossing picture to relativistic orbital dynamics. Ref. \cite{fxnd-r4bs} presents a fully relativistic treatment of EMRIs repeatedly crossing an accretion disk, in which both the orbital dynamics and the disk--secondary interaction are modeled relativistically. Under very similar disk-EMRI configurations, the authors report that the a disk-driven alignment of the secondary's orbital plane, \textit{irrespective} of the initial inclination. Also, they justify that the final eccentricity of a captured object is generically \textit{low}, but that it can undergo a transient growth for initially highly inclined orbits when the secondary is a stellar-mass BH, and that only a modest fraction of nuclear-cluster black holes are captured within a typical AGN-disk lifetime through disk crossings alone. These results are complementary to ours, in several respects. The transient eccentricity growth at large inclination is qualitatively consistent with the non-monotonic, oscillatory behavior we report since Fig. \ref{Fig1-general-dynamics}, where the eccentricity dumps before the system circularizes. In our analysis, this behavior arises within a hybrid framework in which the orbital motion is described by Kerr geodesics, while the disk--secondary interaction is incorporated through an effective interaction model, rather than through a fully relativistic treatment of the disk interaction itself. The qualitative agreement between these two independent approaches therefore provides a useful cross-check of the eccentricity evolution in the highly inclined regime.

We also analyzed Ref. \cite{10.1093/mnras/staa3004}, in which analytic and semi-analytic estimates of capture and alignment timescales for nuclear-cluster orbiters are provided. The latter are assumed to interact with an AGN disk through geometric, drag-dominated crossings, within essentially the same physical picture that we adopt for the disk--secondary interaction. This means that disk crossings can be treated as discrete/quasi-instantaneous events of momentum exchange. The authors likewise find behavior consistent with the two-stage evolution observed in our simulations: (i) an \textit{initial phase} in which the semi-major axis remains approximately constant while the inclination decreases slowly, followed by (ii) a more rapid orbital evolution once some critical relative velocity or inclination is reached. On top of this, the results reported by the authors (together with the subsequent literature on AGN-disk capture and alignment that builds on this framework), is closely related to the physical picture underlying Ref. \cite{Spieksma:2025wex}, on which our Keplerian treatment introduced in Section~\ref{Sec-Preliminaries} is based. Moreover, our results provide a relativistic extension of this picture. At large orbital separations, where relativistic corrections to the orbital dynamics are small, our alignment timescales are consistent, at the order-of-magnitude level, with the non-relativistic estimates obtained in \cite{10.1093/mnras/staa3004} (see Section~\ref{Sec-ResA}). At smaller separations, however, relativistic corrections accumulate and the orbital evolution departs from the corresponding non-relativistic behavior, as discussed in Section~\ref{Sec-ResB}. Thus, the agreement with the previous estimates at large separation provides the expected Newtonian-limit consistency check, while the deviations at smaller separation quantify the regime in which relativistic orbital dynamics becomes essential.

\section{Discussion and outlook}
\label{Sec-Discussion-Outlook}

\subsection{Implications for quasi-periodic eruptions}

In this section we discuss a possible observational application 
of the results presented in this work, namely quasi-periodic 
eruptions (QPEs). These events consist of recurring soft X-ray 
flares observed in time-domain light curves by instruments such as \textit{XMM-Newton} and \textit{Chandra}. To date, a small but 
growing sample of QPE sources has been identified, 
including GSN 069 (a $\sim1\,\rm hr$ soft X-ray flare recurring 
every $\sim9\,\rm hr$) and RX J1301.9, together with several 
additional candidates in the local Universe \cite{2019Natur.573..381M}.

\begin{figure}
    \centering
    \includegraphics[width=0.49\linewidth]{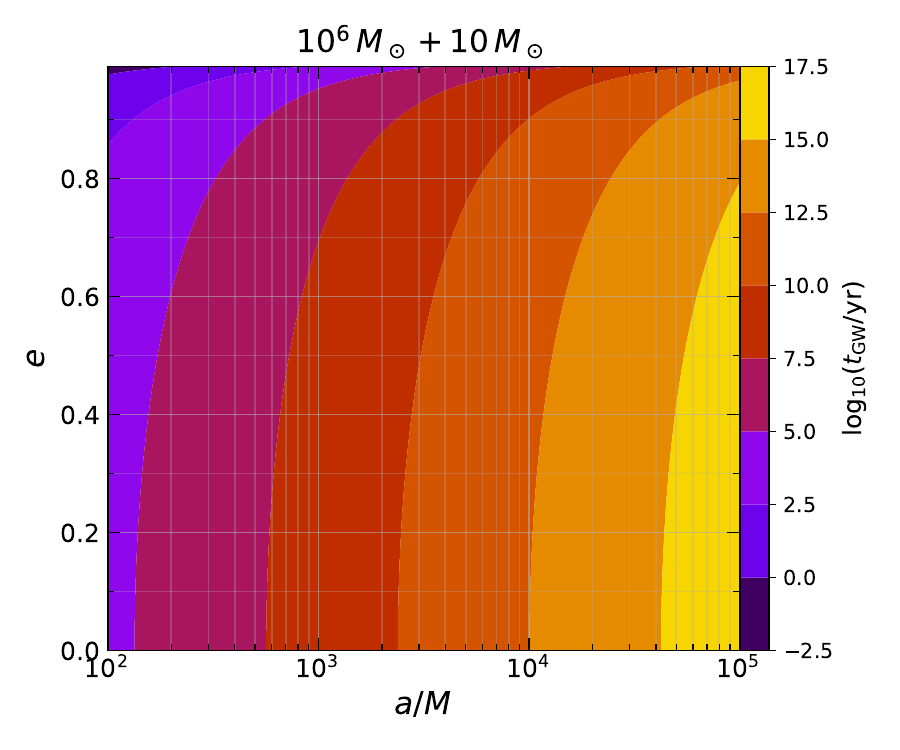}
    \includegraphics[width=0.49\linewidth]{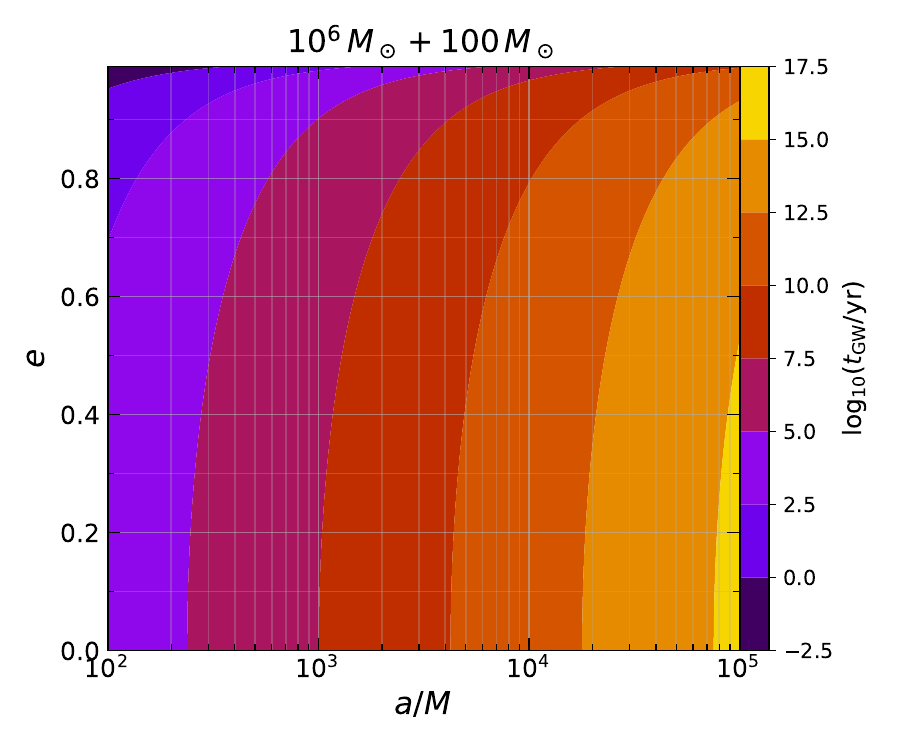}
    \includegraphics[width=0.49\linewidth]{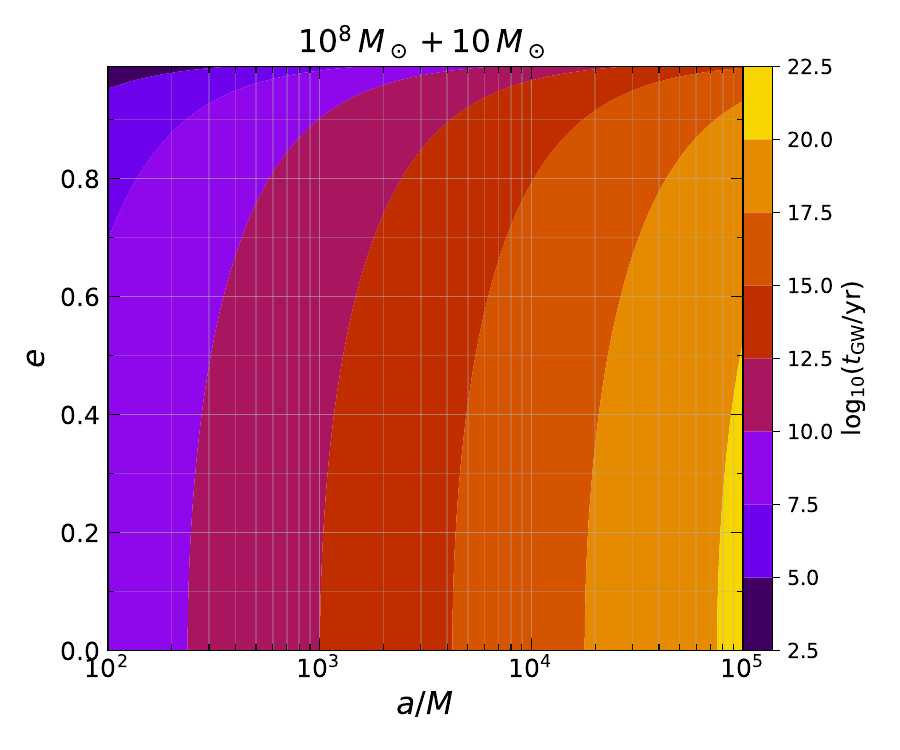}
    \includegraphics[width=0.49\linewidth]{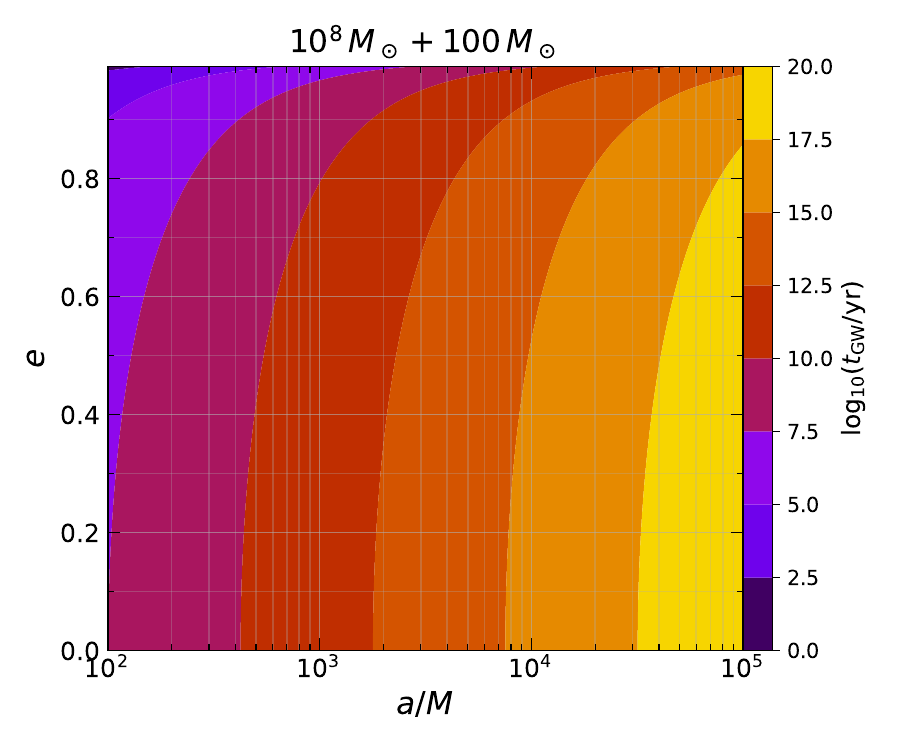}
    \caption{\textit{Gravitational-wave emission.} Timescale of gravitational-wave merging as a function of the initial semi-major axis and eccentricity for an EMRI evolving only through GW energy emission. The top (bottom) panels correspond to a primary mass of $10^6\,M_\odot$ ($10^8\,M_\odot$), while the left (right) panels assume a secondary mass of $10\,M_\odot$ ($100\,M_\odot$). The long merger timescales obtained throughout the region of parameter space relevant to this work shows that GW emission, and therefore the associated 2.5PN contributions, remain subdominant to disk-driven effects.}
    \label{Fig7:GW-timescale}
\end{figure}

The quality of the available observations has motivated a wide 
range of theoretical interpretations. Proposed explanations 
include partial tidal disruption events \cite{Zalamea:2010mv,King:2020jtd,Cufari:2022szx,Nicholl:2024fff} 
and accretion-disk instabilities \cite{2020A&A...641A.167S,2021ApJ...909...82R,2022ApJ...928L..18P}. 
Another possibility involves a compact object orbiting a 
SMBH and repeatedly interacting with a surrounding accretion 
disk \cite{1991MNRAS.250..505S,Linial:2023nqs}. In particular, Ref.~\cite{Linial:2023nqs} considered the case of a star 
orbiting a SMBH, while Ref.~\cite{Franchini:2023bou} showed 
that an EMRI embedded in an accretion disk can reproduce 
several features of observed QPE light curves when the disk 
angular momentum is misaligned with the spin of the 
central BH. Within this picture, X-ray flares are generated 
whenever the orbiting object crosses the disk.

The orbital periods relevant for QPE observations fall 
within the parameter space explored in this work. Indeed, 
for a secondary orbiting a SMBH of mass $M$, the orbital 
period can be estimated from Kepler's law as
\begin{equation}
T \sim 9\,{\rm hr}\,
\left(\frac{M}{10^8\,M_\odot}\right)
\left(\frac{a}{5 M}\right)^{3/2},
\end{equation}
where $a$ denotes the semi-major axis in units of the 
gravitational radius of the primary.

Within the EMRI--disk scenario, an eccentric orbit 
naturally produces two alternating timescales between 
successive bursts. These correspond to disk crossings 
occurring near apocenter and pericenter, yielding a 
longer recurrence time $T_1$ and a shorter one $T_2$, 
respectively. Both timescales scale inversely with 
the local orbital velocity of the secondary. Using 
the Keplerian expressions for the orbital speed at 
the turning points, one finds
\begin{equation}
\frac{T_1}{T_2}\sim \frac{1+e}{1-e}\,,
\end{equation}
where $e$ is the orbital eccentricity. For example, 
an orbit with $e\sim0.6$ predicts $T_1/T_2\sim4$. 
As the orbit circularizes, the two recurrence times 
progressively converge.

This behavior is particularly interesting in light 
of our results. We have shown that repeated disk 
crossings efficiently damp the orbital eccentricity 
while simultaneously driving the system toward 
coplanarity with the disk. Consequently, if disk-driven 
dissipation operates with an efficiency comparable to 
that found in our simulations, QPE systems associated 
with compact objects embedded in AGN disks should evolve 
toward configurations characterized by nearly equal 
recurrence times, $T_1 \sim T_2$. Measurements of 
alternating QPE periods may therefore provide an 
indirect probe of the orbital eccentricity and, 
more generally, of the strength of disk-driven 
dissipation in compact-object systems embedded 
in AGN environments.

\subsection{Conclusions}

In this work we developed a relativistic framework to 
study the secular evolution of EMRIs embedded in an 
accretion disk, with a hydrodynamic prescription for 
repeated disk crossings. This allowed us to consistently 
track the coupled evolution of the eccentricity, 
inclination, and semi-major axis across a wide range 
of orbital configurations.

Our analysis shows that disk--induced dissipation 
generically drives a robust two-stage evolutionary 
scenario, in which orbital-plane alignment occurs on 
(slightly) shorter timescales than eccentricity damping. 
This hierarchy naturally leads to a prolonged phase 
in which systems remain significantly eccentric while 
becoming coplanar with the disk, thereby increasing 
the likelihood of measurable eccentricity in the LISA 
band. Our results suggest a significant dependence of the eccentricity damping rate on the flow 3-velocity, substantially weakening circularization and, in 
some cases, producing damping timescales comparable to 
(or exceeding) the inspiral timescale.

By systematically comparing Keplerian and relativistic 
orbital treatments, we demonstrate that cumulative 
relativistic effects become important even at large 
orbital separations, where one would naively expect 
Keplerian gravity to provide an accurate description. 
These effects are progressively amplified by repeated 
disk crossings, leading to secular deviations in both 
orbital geometry and dissipation rates. In the 
intermediate regime, relativistic corrections 
become dynamically dominant, significantly altering 
migration and damping timescales and highlighting 
the breakdown of purely Keplerian approximations 
well before the strong-field region is reached.

We have also investigated the impact of the accretion 
disk model by comparing the widely used SG prescription 
with a relativistic NT disk constructed numerically 
via the Penna formalism. We find that relativistic disks 
exhibit systematically lower midplane densities and 
larger scale heights in the region $10^2M \lesssim R \lesssim 10^4M$, which leads to a substantially reduced efficiency of 
hydrodynamic drag and, therefore, slower orbital 
evolution. Thus, the choice of the disk model is not 
a secondary detail, but a critical ingredient in 
determining the long-term dynamics of compact objects 
embedded, for instance, in strong AGN disks. In particular, 
Keplerian disk models may systematically overestimate 
the efficiency of orbital capture, alignment, and migration 
in realistic relativistic environments.

Additionally, we have quantified the impact of the BH spin on the 
secular evolution of the system. We find that, despite 
its strong influence on the relativistic spacetime 
structure and on the inner disk region, the spin parameter 
has only a subdominant effect on the overall circularization efficiency, as measured by the number of disk crossings 
required to reach a quasi-circular state. This robustness 
suggests that eccentricity damping is primarily governed 
by cumulative hydrodynamic interactions with the disk 
rather than by spin-induced relativistic corrections.

The results reported in this work open several directions 
for future investigation. A natural extension of the 
present framework consists in developing a fully 
relativistic description of the interaction between the 
secondary and the surrounding disk. While the orbital 
motion is modeled here through exact Kerr geodesics, 
the dissipative processes associated with accretion 
and dynamical friction are still treated using Keplerian 
prescriptions. A self-consistent relativistic treatment 
of both the dynamics and the disk interaction would 
provide a more accurate description of the evolution 
in the strong-field regime.

Another important question concerns the long-term 
astrophysical consequences of disk-driven orbital 
evolution. In particular, it would be interesting to 
investigate whether the eccentricity and inclination 
distributions predicted by our simulations leave 
observable imprints on EMRI populations entering the 
LISA band. Such a study could help establish whether 
gravitational-wave observations retain memory of an 
earlier phase of interaction with an AGN disk.

Finally, the framework developed here may prove useful 
in the context of quasi-periodic eruptions. The orbital 
trajectories generated by our simulations naturally 
determine the sequence and geometry of disk crossings, providing a basis for constructing synthetic flare 
light curves. Comparing such predictions with QPE 
observations may offer a complementary way of 
testing compact-object scenarios in AGN environments.

\begin{acknowledgments}
We would like to thank Thomas Spieksma and Enrico Cannizzaro for very useful comments on an early 
version of this manuscript. We are also grateful to the anonymous referee for constructive and insightful comments that helped improve the manuscript. for M.R was supported by PRIN 2022 grant “GUVIRP - Gravity tests in
the UltraViolet and InfraRed with Pulsar timing”, the
EU Horizon 2020 Research and Innovation Programme
under the Marie Sklodowska-Curie Grant Agreement No.
101007855 and the MUR PRIN Grant No. 2022-Z9X4XS
funded by the European Union (Next Generation EU), and acknowledges current support from Universidad Nacional de C\'ordoba (Argentina). 
K.K. is supported
by NSF grant Nos. AST-2307146, PHY-2513337, PHY090003, and PHY-20043, by NASA grant No. 21-ATP21-
0010, by John Templeton Foundation grant No. 62840,
by the Simons Foundation [MPS-SIP-00001698, E.B.],
by the Simons Foundation International, by Italian Ministry of Foreign Affairs and International Cooperation
grant No. PGR01167, and by the Onassis Foundation
Scholarship (ID: F ZT 041-1/2023-2024).
A.M.~acknowledges financial support from MUR PRIN Grant No.~2022-Z9X4XS.
\end{acknowledgments}

\appendix

\appendix
\section{Conversion between Keplerian orbital parameters and Kerr constants of motion}
\label{app:keplerian-kerr-conversion}

In this appendix, we describe the procedure used to convert the
Keplerian orbital parameters $(a,e,\iota)$, or equivalently
$(a,e,x)$ with $x=\cos\iota$, into the corresponding constants of
motion $(E,L,Q)$ of a bound Kerr geodesic, introduced in Section \ref{Sec-Numerical-Implementation}. This conversion is required
to initialize the geodesic evolution and to reconstruct the orbital
parameters after each disk-crossing interaction.

The radial and polar potentials entering the Kerr geodesic equations
are given in Eqs.~(\ref{R-Kerr}) and~(\ref{R-Theta}) of Section \ref{Sec-Numerical-Implementation}, respectively.
For a bound orbit, the conversion from the orbital parameters to the
Kerr constants is obtained by requiring that the radial motion have
turning points at the apoapsis $r_{\rm a}=a(1+e)$ and periapsis $r_{\rm p}=a(1-e)$, while the polar motion
has a turning point at the maximum orbital latitude $\theta_{\min}=\frac{\pi}{2}-\iota$. We therefore
impose 
\begin{equation}\label{eq:app_turning_conditions}
    R(r_{\rm a})=R(r_{\rm p})=\Theta(\theta_{\min})=0.
\end{equation}
This is equivalent, in terms of $x=\cos\iota$, to
$\cos\theta_{\min}=x$ and
$\sin^2\theta_{\min}=1-x^2$. Then, substituting into the polar
turning-point condition yields
\begin{equation}
Q
=
x^2
\left[
\chi^2(1-E^2)
+
\frac{L^2}{1-x^2}
\right].
\label{eq:app_Q_inclination}
\end{equation}
The remaining two conditions in Eq.~\eqref{eq:app_turning_conditions},
with the radial potential given by Eq.~(\ref{R-Kerr}), constitute now
\textit{two algebraic equations} for $E$ and $L$. Together with
Eq.~\eqref{eq:app_Q_inclination}, they provide a closed system for
the three constants of motion $(E,L,Q)$ corresponding to a specified
set $(a,e,\iota)$.

In practice, we solve this system numerically and select the physical
branch corresponding to the specified bound orbit. In the weak-field
limit, this branch is continuously connected to the corresponding
Newtonian Keplerian orbit. This procedure is equivalent to the
standard conversion between orbital parameters and Kerr constants of
motion described by Schmidt~\cite{Schmidt:2002qk} and employed in the
$(p,e,x)$ parametrization of Fujita and Hikida~\cite{Fujita:2009bp}.
The same conversion underlies the \texttt{KerrGeoPy} implementation
used in our simulations.

\section{Timescale of gravitational-wave emission during the inspiral}
\label{Sec-AppendixA}

In this appendix we analyze the importance of dissipative 
GW effects, justifying why we can safely neglect them 
during the analysis of the binary evolution.

Figure~\ref{Fig7:GW-timescale} shows the merger timescale of an EMRI driven \textit{solely} by GW emission, which we model 
using a 2.5 post-Newtonian (PN) expression, as a function 
of the secondary's initial semi-major axis and eccentricity, following~\cite{Peters:1964zz}. The merger time is computed using the semi-analytic fit of Ref.~\cite{Mandel:2021fra}, which reproduces the exact result to within $3\%$ for eccentricities up to $0.99999$. This comparison allows us to identify the region of the parameter space in which GW energy losses, and therefore the associated 2.5PN dissipative contributions can be safely neglected.

We find that, within the parameters for the EMRIs considered in this work, the evolution is dominated by disk crossings rather than by GW emission. Indeed, for the typical initial conditions considered here, namely $a\sim10^4M$ and eccentricities up to $e\sim0.8$, the inspiral timescale driven by disk interactions is $\tau\lesssim10^6\,\mathrm{yr}$, whereas the merger timescale due \textit{solely} to GW energy loss exceeds $10^9\,\mathrm{yr}$. Gravitational-wave dissipation becomes important only once the binary reaches substantially more compact and eccentric configurations, approximately $a\lesssim10^3M$ and $e\gtrsim0.9$. Since the EMRIs studied here remain well outside this regime during the phase of interest, the contribution of the 2.5PN terms is negligible, thereby justifying their omission from our study. Modeling the late-time evolution once GW dissipation becomes dominant is beyond the scope of the present work and is deferred to future studies.

\section{Eccentricity vs inclination criterion}
\label{Sec-AppendixLast}

In this appendix, we provide a simple geometrical argument justifying why the
eccentricity-based termination criterion used in our simulations
corresponds, in practice, to the onset of the embedded regime in which
discrete disk crossings cease to be well defined.

The argument is geometrical. By the time the eccentricity has decayed
close to $e_{\min}=10^{-3}$, the semi-major axis has also shrunk
substantially relative to its initial value $a_0=10^6\,M$ (see Fig.~\ref{Fig2-dynamics-evo}). Indeed,
dynamical friction and accretion act cumulatively over many crossings,
and by the late stages of the evolution the object is confined to the
innermost portion of its original orbital range. From the top panel of
Fig.~\ref{Fig5-SGvsPenna}, the disk aspect ratio $H/R$ is a strongly \textit{increasing} function
of \textit{decreasing} $R/M$ in the inner region reached during the late stages
of the evolution (both for the SG and the Penna/NT models). The vertical
excursion of a nearly circular, residually inclined orbit of semi-major
axis $a$ and inclination $\iota_{\rm res}$ is
$z_{\max}\simeq a\sin\iota_{\rm res}$; the orbit is effectively embedded
in the disk, in the sense that \textit{it never leaves the region where the
density is non-negligible}, once $z_{\max}\lesssim H(a)$, i.e.
\begin{equation}
\sin\iota_{\rm res} \;\lesssim\; \frac{H(a)}{a}.
\end{equation}
Because $H/a$ grows precisely as $a$ shrinks towards the values reached
once $e\to e_{\min}$, this embedding condition becomes progressively
easier to satisfy as circularization proceeds, \emph{independently of
how small $\iota_{\rm res}$ actually is}. That is to say that the shrinkage
of the orbit that accompanies circularization and the thickening of the
disk at small radii act together to bring the system into the embedded
regime by the time $e$ reaches $e_{\min}$, even though $\iota_{\rm res}$
itself need not be arbitrarily small. This is the geometrical content
behind the statement in footnote~3 that the eccentricity threshold
corresponds to the onset of effective embedding in the disk, rather
than to exact coplanarity.

Finally, we stress that this is a statement about the \emph{late-time,
near-termination} behaviour of the system, and is logically compatible
with the separate statement in Section~\ref{Sec-Numerical-Implementation} that the (early-time)
alignment timescale is an order of magnitude shorter than the
circularization timescale. In this sense, the eccentricity threshold
\textit{does not imply} exact coplanarity; rather, it identifies the point at
which the orbit has become effectively \textit{embedded} in the disk and the
discrete-crossing description ceases to be applicable.

\bibliography{crossings}

\providecommand{\noopsort}[1]{}\providecommand{\singleletter}[1]{#1}%
\begin{thebibliography}{53}%
\makeatletter
\providecommand \@ifxundefined [1]{%
 \@ifx{#1\undefined}
}%
\providecommand \@ifnum [1]{%
 \ifnum #1\expandafter \@firstoftwo
 \else \expandafter \@secondoftwo
 \fi
}%
\providecommand \@ifx [1]{%
 \ifx #1\expandafter \@firstoftwo
 \else \expandafter \@secondoftwo
 \fi
}%
\providecommand \natexlab [1]{#1}%
\providecommand \enquote  [1]{``#1''}%
\providecommand \bibnamefont  [1]{#1}%
\providecommand \bibfnamefont [1]{#1}%
\providecommand \citenamefont [1]{#1}%
\providecommand \href@noop [0]{\@secondoftwo}%
\providecommand \href [0]{\begingroup \@sanitize@url \@href}%
\providecommand \@href[1]{\@@startlink{#1}\@@href}%
\providecommand \@@href[1]{\endgroup#1\@@endlink}%
\providecommand \@sanitize@url [0]{\catcode `\\12\catcode `\$12\catcode `\&12\catcode `\#12\catcode `\^12\catcode `\_12\catcode `\%12\relax}%
\providecommand \@@startlink[1]{}%
\providecommand \@@endlink[0]{}%
\providecommand \url  [0]{\begingroup\@sanitize@url \@url }%
\providecommand \@url [1]{\endgroup\@href {#1}{\urlprefix }}%
\providecommand \urlprefix  [0]{URL }%
\providecommand \Eprint [0]{\href }%
\providecommand \doibase [0]{https://doi.org/}%
\providecommand \selectlanguage [0]{\@gobble}%
\providecommand \bibinfo  [0]{\@secondoftwo}%
\providecommand \bibfield  [0]{\@secondoftwo}%
\providecommand \translation [1]{[#1]}%
\providecommand \BibitemOpen [0]{}%
\providecommand \bibitemStop [0]{}%
\providecommand \bibitemNoStop [0]{.\EOS\space}%
\providecommand \EOS [0]{\spacefactor3000\relax}%
\providecommand \BibitemShut  [1]{\csname bibitem#1\endcsname}%
\let\auto@bib@innerbib\@empty
\bibitem [{\citenamefont {Cardoso}\ and\ \citenamefont {Pani}(2019)}]{Cardoso:2019rvt}%
  \BibitemOpen
  \bibfield  {author} {\bibinfo {author} {\bibfnamefont {V.}~\bibnamefont {Cardoso}}\ and\ \bibinfo {author} {\bibfnamefont {P.}~\bibnamefont {Pani}},\ }\href {https://doi.org/10.1007/s41114-019-0020-4} {\bibfield  {journal} {\bibinfo  {journal} {Living Rev. Rel.}\ }\textbf {\bibinfo {volume} {22}},\ \bibinfo {pages} {4} (\bibinfo {year} {2019})},\ \Eprint {https://arxiv.org/abs/1904.05363} {arXiv:1904.05363 [gr-qc]} \BibitemShut {NoStop}%
\bibitem [{\citenamefont {Barausse}\ \emph {et~al.}(2014)\citenamefont {Barausse}, \citenamefont {Cardoso},\ and\ \citenamefont {Pani}}]{Barausse2014_EnvEffects}%
  \BibitemOpen
  \bibfield  {author} {\bibinfo {author} {\bibfnamefont {E.}~\bibnamefont {Barausse}}, \bibinfo {author} {\bibfnamefont {V.}~\bibnamefont {Cardoso}},\ and\ \bibinfo {author} {\bibfnamefont {P.}~\bibnamefont {Pani}},\ }\href {https://doi.org/10.1103/PhysRevD.89.104059} {\bibfield  {journal} {\bibinfo  {journal} {Physical Review D}\ }\textbf {\bibinfo {volume} {89}},\ \bibinfo {pages} {104059} (\bibinfo {year} {2014})},\ \Eprint {https://arxiv.org/abs/1404.7149} {arXiv:1404.7149 [astro-ph.CO]} \BibitemShut {NoStop}%
\bibitem [{\citenamefont {Seoane}\ \emph {et~al.}(2023)\citenamefont {Seoane} \emph {et~al.}}]{LISA:2022yao}%
  \BibitemOpen
  \bibfield  {author} {\bibinfo {author} {\bibfnamefont {P.~A.}\ \bibnamefont {Seoane}} \emph {et~al.} (\bibinfo {collaboration} {LISA}),\ }\href {https://doi.org/10.1007/s41114-022-00041-y} {\bibfield  {journal} {\bibinfo  {journal} {Living Rev. Rel.}\ }\textbf {\bibinfo {volume} {26}},\ \bibinfo {pages} {2} (\bibinfo {year} {2023})},\ \Eprint {https://arxiv.org/abs/2203.06016} {arXiv:2203.06016 [gr-qc]} \BibitemShut {NoStop}%
\bibitem [{\citenamefont {Arun}\ \emph {et~al.}(2022)\citenamefont {Arun} \emph {et~al.}}]{LISA:2022kgy}%
  \BibitemOpen
  \bibfield  {author} {\bibinfo {author} {\bibfnamefont {K.~G.}\ \bibnamefont {Arun}} \emph {et~al.} (\bibinfo {collaboration} {LISA}),\ }\href {https://doi.org/10.1007/s41114-022-00036-9} {\bibfield  {journal} {\bibinfo  {journal} {Living Rev. Rel.}\ }\textbf {\bibinfo {volume} {25}},\ \bibinfo {pages} {4} (\bibinfo {year} {2022})},\ \Eprint {https://arxiv.org/abs/2205.01597} {arXiv:2205.01597 [gr-qc]} \BibitemShut {NoStop}%
\bibitem [{\citenamefont {Broekgaarden}\ \emph {et~al.}(2024)\citenamefont {Broekgaarden}, \citenamefont {Banagiri},\ and\ \citenamefont {Payne}}]{Broekgaarden:2023rta}%
  \BibitemOpen
  \bibfield  {author} {\bibinfo {author} {\bibfnamefont {F.~S.}\ \bibnamefont {Broekgaarden}}, \bibinfo {author} {\bibfnamefont {S.}~\bibnamefont {Banagiri}},\ and\ \bibinfo {author} {\bibfnamefont {E.}~\bibnamefont {Payne}},\ }\href {https://doi.org/10.3847/1538-4357/ad4709} {\bibfield  {journal} {\bibinfo  {journal} {Astrophys. J.}\ }\textbf {\bibinfo {volume} {969}},\ \bibinfo {pages} {108} (\bibinfo {year} {2024})},\ \Eprint {https://arxiv.org/abs/2303.17628} {arXiv:2303.17628 [astro-ph.HE]} \BibitemShut {NoStop}%
\bibitem [{\citenamefont {Toubiana}\ \emph {et~al.}(2021)\citenamefont {Toubiana}, \citenamefont {Sberna}, \citenamefont {Caputo}, \citenamefont {Cusin}, \citenamefont {Marsat}, \citenamefont {Jani}, \citenamefont {Babak}, \citenamefont {Barausse}, \citenamefont {Caprini}, \citenamefont {Pani}, \citenamefont {Sesana},\ and\ \citenamefont {Tamanini}}]{PhysRevLett.126.101105}%
  \BibitemOpen
  \bibfield  {author} {\bibinfo {author} {\bibfnamefont {A.}~\bibnamefont {Toubiana}}, \bibinfo {author} {\bibfnamefont {L.}~\bibnamefont {Sberna}}, \bibinfo {author} {\bibfnamefont {A.}~\bibnamefont {Caputo}}, \bibinfo {author} {\bibfnamefont {G.}~\bibnamefont {Cusin}}, \bibinfo {author} {\bibfnamefont {S.}~\bibnamefont {Marsat}}, \bibinfo {author} {\bibfnamefont {K.}~\bibnamefont {Jani}}, \bibinfo {author} {\bibfnamefont {S.}~\bibnamefont {Babak}}, \bibinfo {author} {\bibfnamefont {E.}~\bibnamefont {Barausse}}, \bibinfo {author} {\bibfnamefont {C.}~\bibnamefont {Caprini}}, \bibinfo {author} {\bibfnamefont {P.}~\bibnamefont {Pani}}, \bibinfo {author} {\bibfnamefont {A.}~\bibnamefont {Sesana}},\ and\ \bibinfo {author} {\bibfnamefont {N.}~\bibnamefont {Tamanini}},\ }\href {https://doi.org/10.1103/PhysRevLett.126.101105} {\bibfield  {journal} {\bibinfo  {journal} {Phys. Rev. Lett.}\ }\textbf {\bibinfo {volume} {126}},\ \bibinfo {pages} {101105} (\bibinfo {year} {2021})}\BibitemShut {NoStop}%
\bibitem [{\citenamefont {Cardoso}\ and\ \citenamefont {Maselli}(2020)}]{Cardoso:2019rou}%
  \BibitemOpen
  \bibfield  {author} {\bibinfo {author} {\bibfnamefont {V.}~\bibnamefont {Cardoso}}\ and\ \bibinfo {author} {\bibfnamefont {A.}~\bibnamefont {Maselli}},\ }\href {https://doi.org/10.1051/0004-6361/202037654} {\bibfield  {journal} {\bibinfo  {journal} {Astron. Astrophys.}\ }\textbf {\bibinfo {volume} {644}},\ \bibinfo {pages} {A147} (\bibinfo {year} {2020})},\ \Eprint {https://arxiv.org/abs/1909.05870} {arXiv:1909.05870 [astro-ph.HE]} \BibitemShut {NoStop}%
\bibitem [{\citenamefont {Zwick}\ \emph {et~al.}(2024)\citenamefont {Zwick}, \citenamefont {Capelo},\ and\ \citenamefont {Mayer}}]{Zwick2024_NovelEnvPRD}%
  \BibitemOpen
  \bibfield  {author} {\bibinfo {author} {\bibfnamefont {L.}~\bibnamefont {Zwick}}, \bibinfo {author} {\bibfnamefont {P.~R.}\ \bibnamefont {Capelo}},\ and\ \bibinfo {author} {\bibfnamefont {L.}~\bibnamefont {Mayer}},\ }\href {https://doi.org/10.1103/PhysRevD.110.103005} {\bibfield  {journal} {\bibinfo  {journal} {Physical Review D}\ }\textbf {\bibinfo {volume} {110}},\ \bibinfo {pages} {103005} (\bibinfo {year} {2024})},\ \Eprint {https://arxiv.org/abs/2405.05698} {arXiv:2405.05698 [astro-ph.HE]} \BibitemShut {NoStop}%
\bibitem [{\citenamefont {Amaro-Seoane}(2018)}]{Amaro-Seoane:2012lgq}%
  \BibitemOpen
  \bibfield  {author} {\bibinfo {author} {\bibfnamefont {P.}~\bibnamefont {Amaro-Seoane}},\ }\href {https://doi.org/10.1007/s41114-018-0013-8} {\bibfield  {journal} {\bibinfo  {journal} {Living Rev. Rel.}\ }\textbf {\bibinfo {volume} {21}},\ \bibinfo {pages} {4} (\bibinfo {year} {2018})},\ \Eprint {https://arxiv.org/abs/1205.5240} {arXiv:1205.5240 [astro-ph.CO]} \BibitemShut {NoStop}%
\bibitem [{\citenamefont {Barack}\ \emph {et~al.}(2019)\citenamefont {Barack} \emph {et~al.}}]{Barack:2018yly}%
  \BibitemOpen
  \bibfield  {author} {\bibinfo {author} {\bibfnamefont {L.}~\bibnamefont {Barack}} \emph {et~al.},\ }\href {https://doi.org/10.1088/1361-6382/ab0587} {\bibfield  {journal} {\bibinfo  {journal} {Class. Quant. Grav.}\ }\textbf {\bibinfo {volume} {36}},\ \bibinfo {pages} {143001} (\bibinfo {year} {2019})},\ \Eprint {https://arxiv.org/abs/1806.05195} {arXiv:1806.05195 [gr-qc]} \BibitemShut {NoStop}%
\bibitem [{\citenamefont {Barausse}\ and\ \citenamefont {Rezzolla}(2008)}]{BarausseRezzolla2008}%
  \BibitemOpen
  \bibfield  {author} {\bibinfo {author} {\bibfnamefont {E.}~\bibnamefont {Barausse}}\ and\ \bibinfo {author} {\bibfnamefont {L.}~\bibnamefont {Rezzolla}},\ }\href {https://doi.org/10.1103/PhysRevD.77.104027} {\bibfield  {journal} {\bibinfo  {journal} {Physical Review D}\ }\textbf {\bibinfo {volume} {77}} (\bibinfo {year} {2008})}\BibitemShut {NoStop}%
\bibitem [{\citenamefont {Nouri}\ and\ \citenamefont {Janiuk}(2024)}]{ViscousTorqueGN2024}%
  \BibitemOpen
  \bibfield  {author} {\bibinfo {author} {\bibfnamefont {F.~H.}\ \bibnamefont {Nouri}}\ and\ \bibinfo {author} {\bibfnamefont {A.}~\bibnamefont {Janiuk}},\ }\bibfield  {journal} {\bibinfo  {journal} {Astronomy I\& Astrophysics}\ }\textbf {\bibinfo {volume} {669}},\ \href {https://doi.org/10.1051/0004-6361/20248796-23} {10.1051/0004-6361/20248796-23} (\bibinfo {year} {2024})\BibitemShut {NoStop}%
\bibitem [{\citenamefont {Cardoso}\ \emph {et~al.}(2022)\citenamefont {Cardoso}, \citenamefont {Destounis}, \citenamefont {Duque}, \citenamefont {Macedo},\ and\ \citenamefont {Maselli}}]{PhysRevLett.129.241103}%
  \BibitemOpen
  \bibfield  {author} {\bibinfo {author} {\bibfnamefont {V.}~\bibnamefont {Cardoso}}, \bibinfo {author} {\bibfnamefont {K.}~\bibnamefont {Destounis}}, \bibinfo {author} {\bibfnamefont {F.}~\bibnamefont {Duque}}, \bibinfo {author} {\bibfnamefont {R.~P.}\ \bibnamefont {Macedo}},\ and\ \bibinfo {author} {\bibfnamefont {A.}~\bibnamefont {Maselli}},\ }\href {https://doi.org/10.1103/PhysRevLett.129.241103} {\bibfield  {journal} {\bibinfo  {journal} {Phys. Rev. Lett.}\ }\textbf {\bibinfo {volume} {129}},\ \bibinfo {pages} {241103} (\bibinfo {year} {2022})}\BibitemShut {NoStop}%
\bibitem [{\citenamefont {Grishin}\ \emph {et~al.}(2024)\citenamefont {Grishin}, \citenamefont {Gilbaum},\ and\ \citenamefont {Stone}}]{Grishin:2023riv}%
  \BibitemOpen
  \bibfield  {author} {\bibinfo {author} {\bibfnamefont {E.}~\bibnamefont {Grishin}}, \bibinfo {author} {\bibfnamefont {S.}~\bibnamefont {Gilbaum}},\ and\ \bibinfo {author} {\bibfnamefont {N.~C.}\ \bibnamefont {Stone}},\ }\href {https://doi.org/10.1093/mnras/stae828} {\bibfield  {journal} {\bibinfo  {journal} {Mon. Not. Roy. Astron. Soc.}\ }\textbf {\bibinfo {volume} {530}},\ \bibinfo {pages} {2114} (\bibinfo {year} {2024})},\ \bibinfo {note} {[Erratum: Mon.Not.Roy.Astron.Soc. 535, 1693--1694 (2024)]},\ \Eprint {https://arxiv.org/abs/2307.07546} {arXiv:2307.07546 [astro-ph.HE]} \BibitemShut {NoStop}%
\bibitem [{\citenamefont {Sedda}\ \emph {et~al.}(2023)\citenamefont {Sedda}, \citenamefont {Naoz},\ and\ \citenamefont {Kocsis}}]{Sedda:2023big}%
  \BibitemOpen
  \bibfield  {author} {\bibinfo {author} {\bibfnamefont {M.~A.}\ \bibnamefont {Sedda}}, \bibinfo {author} {\bibfnamefont {S.}~\bibnamefont {Naoz}},\ and\ \bibinfo {author} {\bibfnamefont {B.}~\bibnamefont {Kocsis}},\ }\href {https://doi.org/10.3390/universe9030138} {\bibfield  {journal} {\bibinfo  {journal} {Universe}\ }\textbf {\bibinfo {volume} {9}},\ \bibinfo {pages} {138} (\bibinfo {year} {2023})},\ \Eprint {https://arxiv.org/abs/2302.14071} {arXiv:2302.14071 [astro-ph.GA]} \BibitemShut {NoStop}%
\bibitem [{\citenamefont {Zeng}\ and\ \citenamefont {Pan}(2026)}]{fxnd-r4bs}%
  \BibitemOpen
  \bibfield  {author} {\bibinfo {author} {\bibfnamefont {Y.}~\bibnamefont {Zeng}}\ and\ \bibinfo {author} {\bibfnamefont {Z.}~\bibnamefont {Pan}},\ }\href {https://doi.org/10.1103/fxnd-r4bs} {\bibfield  {journal} {\bibinfo  {journal} {Phys. Rev. D}\ }\textbf {\bibinfo {volume} {113}},\ \bibinfo {pages} {123002} (\bibinfo {year} {2026})}\BibitemShut {NoStop}%
\bibitem [{\citenamefont {Fabj}\ \emph {et~al.}(2020)\citenamefont {Fabj}, \citenamefont {Nasim}, \citenamefont {Caban}, \citenamefont {Ford}, \citenamefont {McKernan},\ and\ \citenamefont {Bellovary}}]{10.1093/mnras/staa3004}%
  \BibitemOpen
  \bibfield  {author} {\bibinfo {author} {\bibfnamefont {G.}~\bibnamefont {Fabj}}, \bibinfo {author} {\bibfnamefont {S.~S.}\ \bibnamefont {Nasim}}, \bibinfo {author} {\bibfnamefont {F.}~\bibnamefont {Caban}}, \bibinfo {author} {\bibfnamefont {K.~E.~S.}\ \bibnamefont {Ford}}, \bibinfo {author} {\bibfnamefont {B.}~\bibnamefont {McKernan}},\ and\ \bibinfo {author} {\bibfnamefont {J.~M.}\ \bibnamefont {Bellovary}},\ }\href {https://doi.org/10.1093/mnras/staa3004} {\bibfield  {journal} {\bibinfo  {journal} {Monthly Notices of the Royal Astronomical Society}\ }\textbf {\bibinfo {volume} {499}},\ \bibinfo {pages} {2608} (\bibinfo {year} {2020})},\ \Eprint {https://arxiv.org/abs/https://academic.oup.com/mnras/article-pdf/499/2/2608/33980336/staa3004.pdf} {https://academic.oup.com/mnras/article-pdf/499/2/2608/33980336/staa3004.pdf} \BibitemShut {NoStop}%
\bibitem [{\citenamefont {Linial}\ and\ \citenamefont {Metzger}(2023)}]{Linial:2023nqs}%
  \BibitemOpen
  \bibfield  {author} {\bibinfo {author} {\bibfnamefont {I.}~\bibnamefont {Linial}}\ and\ \bibinfo {author} {\bibfnamefont {B.~D.}\ \bibnamefont {Metzger}},\ }\href {https://doi.org/10.3847/1538-4357/acf65b} {\bibfield  {journal} {\bibinfo  {journal} {Astrophys. J.}\ }\textbf {\bibinfo {volume} {957}},\ \bibinfo {pages} {34} (\bibinfo {year} {2023})},\ \Eprint {https://arxiv.org/abs/2303.16231} {arXiv:2303.16231 [astro-ph.HE]} \BibitemShut {NoStop}%
\bibitem [{\citenamefont {Franchini}\ \emph {et~al.}(2023)\citenamefont {Franchini}, \citenamefont {Bonetti}, \citenamefont {Lupi}, \citenamefont {Miniutti}, \citenamefont {Bortolas}, \citenamefont {Giustini}, \citenamefont {Dotti}, \citenamefont {Sesana}, \citenamefont {Arcodia},\ and\ \citenamefont {Ryu}}]{Franchini:2023bou}%
  \BibitemOpen
  \bibfield  {author} {\bibinfo {author} {\bibfnamefont {A.}~\bibnamefont {Franchini}}, \bibinfo {author} {\bibfnamefont {M.}~\bibnamefont {Bonetti}}, \bibinfo {author} {\bibfnamefont {A.}~\bibnamefont {Lupi}}, \bibinfo {author} {\bibfnamefont {G.}~\bibnamefont {Miniutti}}, \bibinfo {author} {\bibfnamefont {E.}~\bibnamefont {Bortolas}}, \bibinfo {author} {\bibfnamefont {M.}~\bibnamefont {Giustini}}, \bibinfo {author} {\bibfnamefont {M.}~\bibnamefont {Dotti}}, \bibinfo {author} {\bibfnamefont {A.}~\bibnamefont {Sesana}}, \bibinfo {author} {\bibfnamefont {R.}~\bibnamefont {Arcodia}},\ and\ \bibinfo {author} {\bibfnamefont {T.}~\bibnamefont {Ryu}},\ }\href {https://doi.org/10.1051/0004-6361/202346565} {\bibfield  {journal} {\bibinfo  {journal} {Astron. Astrophys.}\ }\textbf {\bibinfo {volume} {675}},\ \bibinfo {pages} {A100} (\bibinfo {year} {2023})},\ \Eprint {https://arxiv.org/abs/2304.00775} {arXiv:2304.00775 [astro-ph.HE]} \BibitemShut {NoStop}%
\bibitem [{\citenamefont {Spieksma}\ and\ \citenamefont {Cannizzaro}(2026)}]{Spieksma:2025wex}%
  \BibitemOpen
  \bibfield  {author} {\bibinfo {author} {\bibfnamefont {T.~F.~M.}\ \bibnamefont {Spieksma}}\ and\ \bibinfo {author} {\bibfnamefont {E.}~\bibnamefont {Cannizzaro}},\ }\href {https://doi.org/10.1093/mnras/stag021} {\bibfield  {journal} {\bibinfo  {journal} {Mon. Not. Roy. Astron. Soc.}\ }\textbf {\bibinfo {volume} {546}},\ \bibinfo {pages} {stag021} (\bibinfo {year} {2026})},\ \Eprint {https://arxiv.org/abs/2504.08033} {arXiv:2504.08033 [astro-ph.GA]} \BibitemShut {NoStop}%
\bibitem [{\citenamefont {Pringle}(1981)}]{Pringle:1981ds}%
  \BibitemOpen
  \bibfield  {author} {\bibinfo {author} {\bibfnamefont {J.~E.}\ \bibnamefont {Pringle}},\ }\href {https://doi.org/10.1146/annurev.aa.19.090181.001033} {\bibfield  {journal} {\bibinfo  {journal} {Ann. Rev. Astron. Astrophys.}\ }\textbf {\bibinfo {volume} {19}},\ \bibinfo {pages} {137} (\bibinfo {year} {1981})}\BibitemShut {NoStop}%
\bibitem [{\citenamefont {Shakura}\ and\ \citenamefont {Sunyaev}(1973)}]{Shakura:1972te}%
  \BibitemOpen
  \bibfield  {author} {\bibinfo {author} {\bibfnamefont {N.~I.}\ \bibnamefont {Shakura}}\ and\ \bibinfo {author} {\bibfnamefont {R.~A.}\ \bibnamefont {Sunyaev}},\ }\href@noop {} {\bibfield  {journal} {\bibinfo  {journal} {Astron. Astrophys.}\ }\textbf {\bibinfo {volume} {24}},\ \bibinfo {pages} {337} (\bibinfo {year} {1973})}\BibitemShut {NoStop}%
\bibitem [{\citenamefont {Shakura}\ and\ \citenamefont {Sunyaev}(1976)}]{Shakura:1976xk}%
  \BibitemOpen
  \bibfield  {author} {\bibinfo {author} {\bibfnamefont {N.~I.}\ \bibnamefont {Shakura}}\ and\ \bibinfo {author} {\bibfnamefont {R.~A.}\ \bibnamefont {Sunyaev}},\ }\href@noop {} {\bibfield  {journal} {\bibinfo  {journal} {Mon. Not. Roy. Astron. Soc.}\ }\textbf {\bibinfo {volume} {175}},\ \bibinfo {pages} {613} (\bibinfo {year} {1976})}\BibitemShut {NoStop}%
\bibitem [{\citenamefont {Fragile}\ and\ \citenamefont {Liska}(2025)}]{Fragile:2024osq}%
  \BibitemOpen
  \bibfield  {author} {\bibinfo {author} {\bibfnamefont {P.~C.}\ \bibnamefont {Fragile}}\ and\ \bibinfo {author} {\bibfnamefont {M.}~\bibnamefont {Liska}},\ }\bibinfo {title} {{Tilted Accretion Disks}}\ (\bibinfo {year} {2025})\ \Eprint {https://arxiv.org/abs/2404.10052} {arXiv:2404.10052 [astro-ph.HE]} \BibitemShut {NoStop}%
\bibitem [{\citenamefont {Dihingia}\ and\ \citenamefont {Fendt}(2025)}]{Dihingia:2024tqr}%
  \BibitemOpen
  \bibfield  {author} {\bibinfo {author} {\bibfnamefont {I.~K.}\ \bibnamefont {Dihingia}}\ and\ \bibinfo {author} {\bibfnamefont {C.}~\bibnamefont {Fendt}},\ }\bibinfo {title} {{Thin Accretion Disks in~GR-MHD Simulations}}\ (\bibinfo {year} {2025})\ \Eprint {https://arxiv.org/abs/2404.06140} {arXiv:2404.06140 [astro-ph.HE]} \BibitemShut {NoStop}%
\bibitem [{\citenamefont {Abramowicz}\ and\ \citenamefont {Fragile}(2013)}]{Abramowicz:2011xu}%
  \BibitemOpen
  \bibfield  {author} {\bibinfo {author} {\bibfnamefont {M.~A.}\ \bibnamefont {Abramowicz}}\ and\ \bibinfo {author} {\bibfnamefont {P.~C.}\ \bibnamefont {Fragile}},\ }\href {https://doi.org/10.12942/lrr-2013-1} {\bibfield  {journal} {\bibinfo  {journal} {Living Rev. Rel.}\ }\textbf {\bibinfo {volume} {16}},\ \bibinfo {pages} {1} (\bibinfo {year} {2013})},\ \Eprint {https://arxiv.org/abs/1104.5499} {arXiv:1104.5499 [astro-ph.HE]} \BibitemShut {NoStop}%
\bibitem [{\citenamefont {Penna}\ \emph {et~al.}(2010)\citenamefont {Penna}, \citenamefont {McKinney}, \citenamefont {Narayan}, \citenamefont {Tchekhovskoy}, \citenamefont {Shafee},\ and\ \citenamefont {McClintock}}]{10.1111/j.1365-2966.2010.17170.x}%
  \BibitemOpen
  \bibfield  {author} {\bibinfo {author} {\bibfnamefont {R.~F.}\ \bibnamefont {Penna}}, \bibinfo {author} {\bibfnamefont {J.~C.}\ \bibnamefont {McKinney}}, \bibinfo {author} {\bibfnamefont {R.}~\bibnamefont {Narayan}}, \bibinfo {author} {\bibfnamefont {A.}~\bibnamefont {Tchekhovskoy}}, \bibinfo {author} {\bibfnamefont {R.}~\bibnamefont {Shafee}},\ and\ \bibinfo {author} {\bibfnamefont {J.~E.}\ \bibnamefont {McClintock}},\ }\href {https://doi.org/10.1111/j.1365-2966.2010.17170.x} {\bibfield  {journal} {\bibinfo  {journal} {Monthly Notices of the Royal Astronomical Society}\ }\textbf {\bibinfo {volume} {408}},\ \bibinfo {pages} {752} (\bibinfo {year} {2010})},\ \Eprint {https://arxiv.org/abs/https://academic.oup.com/mnras/article-pdf/408/2/752/18433142/mnras0408-0752.pdf} {https://academic.oup.com/mnras/article-pdf/408/2/752/18433142/mnras0408-0752.pdf} \BibitemShut {NoStop}%
\bibitem [{\citenamefont {Toomre}(1964)}]{Toomre:1964zx}%
  \BibitemOpen
  \bibfield  {author} {\bibinfo {author} {\bibfnamefont {A.}~\bibnamefont {Toomre}},\ }\href {https://doi.org/10.1086/147861} {\bibfield  {journal} {\bibinfo  {journal} {Astrophys. J.}\ }\textbf {\bibinfo {volume} {139}},\ \bibinfo {pages} {1217} (\bibinfo {year} {1964})}\BibitemShut {NoStop}%
\bibitem [{\citenamefont {Kratter}\ and\ \citenamefont {Lodato}(2016)}]{kratter2016gravitational}%
  \BibitemOpen
  \bibfield  {author} {\bibinfo {author} {\bibfnamefont {K.}~\bibnamefont {Kratter}}\ and\ \bibinfo {author} {\bibfnamefont {G.}~\bibnamefont {Lodato}},\ }\href@noop {} {\bibfield  {journal} {\bibinfo  {journal} {Annual Review of Astronomy and Astrophysics}\ }\textbf {\bibinfo {volume} {54}},\ \bibinfo {pages} {271} (\bibinfo {year} {2016})}\BibitemShut {NoStop}%
\bibitem [{\citenamefont {Page}\ and\ \citenamefont {Thorne}(1974)}]{Page:1974he}%
  \BibitemOpen
  \bibfield  {author} {\bibinfo {author} {\bibfnamefont {D.~N.}\ \bibnamefont {Page}}\ and\ \bibinfo {author} {\bibfnamefont {K.~S.}\ \bibnamefont {Thorne}},\ }\href {https://doi.org/10.1086/152990} {\bibfield  {journal} {\bibinfo  {journal} {Astrophys. J.}\ }\textbf {\bibinfo {volume} {191}},\ \bibinfo {pages} {499} (\bibinfo {year} {1974})}\BibitemShut {NoStop}%
\bibitem [{\citenamefont {Agol}\ and\ \citenamefont {Krolik}(2000)}]{Agol:1999dn}%
  \BibitemOpen
  \bibfield  {author} {\bibinfo {author} {\bibfnamefont {E.}~\bibnamefont {Agol}}\ and\ \bibinfo {author} {\bibfnamefont {J.}~\bibnamefont {Krolik}},\ }\href {https://doi.org/10.1086/308177} {\bibfield  {journal} {\bibinfo  {journal} {Astrophys. J.}\ }\textbf {\bibinfo {volume} {528}},\ \bibinfo {pages} {161} (\bibinfo {year} {2000})},\ \Eprint {https://arxiv.org/abs/astro-ph/9908049} {arXiv:astro-ph/9908049} \BibitemShut {NoStop}%
\bibitem [{\citenamefont {Krolik}\ and\ \citenamefont {Hawley}(2002)}]{Krolik:2002ae}%
  \BibitemOpen
  \bibfield  {author} {\bibinfo {author} {\bibfnamefont {J.~H.}\ \bibnamefont {Krolik}}\ and\ \bibinfo {author} {\bibfnamefont {J.~F.}\ \bibnamefont {Hawley}},\ }\href {https://doi.org/10.1086/340760} {\bibfield  {journal} {\bibinfo  {journal} {Astrophys. J.}\ }\textbf {\bibinfo {volume} {573}},\ \bibinfo {pages} {754} (\bibinfo {year} {2002})},\ \Eprint {https://arxiv.org/abs/astro-ph/0203289} {arXiv:astro-ph/0203289} \BibitemShut {NoStop}%
\bibitem [{\citenamefont {Kley}\ and\ \citenamefont {Nelson}(2012)}]{Kley:2012ue}%
  \BibitemOpen
  \bibfield  {author} {\bibinfo {author} {\bibfnamefont {W.}~\bibnamefont {Kley}}\ and\ \bibinfo {author} {\bibfnamefont {R.~P.}\ \bibnamefont {Nelson}},\ }\href {https://doi.org/10.1146/annurev-astro-081811-125523} {\bibfield  {journal} {\bibinfo  {journal} {Ann. Rev. Astron. Astrophys.}\ }\textbf {\bibinfo {volume} {50}},\ \bibinfo {pages} {211} (\bibinfo {year} {2012})},\ \Eprint {https://arxiv.org/abs/1203.1184} {arXiv:1203.1184 [astro-ph.EP]} \BibitemShut {NoStop}%
\bibitem [{\citenamefont {DeLaurentiis}\ \emph {et~al.}(2023)\citenamefont {DeLaurentiis}, \citenamefont {Epstein-Martin},\ and\ \citenamefont {Haiman}}]{DeLaurentiis:2022qjq}%
  \BibitemOpen
  \bibfield  {author} {\bibinfo {author} {\bibfnamefont {S.}~\bibnamefont {DeLaurentiis}}, \bibinfo {author} {\bibfnamefont {M.}~\bibnamefont {Epstein-Martin}},\ and\ \bibinfo {author} {\bibfnamefont {Z.}~\bibnamefont {Haiman}},\ }\href {https://doi.org/10.1093/mnras/stad1412} {\bibfield  {journal} {\bibinfo  {journal} {Mon. Not. Roy. Astron. Soc.}\ }\textbf {\bibinfo {volume} {523}},\ \bibinfo {pages} {1126} (\bibinfo {year} {2023})},\ \Eprint {https://arxiv.org/abs/2212.02650} {arXiv:2212.02650 [astro-ph.HE]} \BibitemShut {NoStop}%
\bibitem [{\citenamefont {Ostriker}(1999)}]{Ostriker:1998fa}%
  \BibitemOpen
  \bibfield  {author} {\bibinfo {author} {\bibfnamefont {E.~C.}\ \bibnamefont {Ostriker}},\ }\href {https://doi.org/10.1086/306858} {\bibfield  {journal} {\bibinfo  {journal} {Astrophys. J.}\ }\textbf {\bibinfo {volume} {513}},\ \bibinfo {pages} {252} (\bibinfo {year} {1999})},\ \Eprint {https://arxiv.org/abs/astro-ph/9810324} {arXiv:astro-ph/9810324} \BibitemShut {NoStop}%
\bibitem [{\citenamefont {Tagawa}\ \emph {et~al.}(2020)\citenamefont {Tagawa}, \citenamefont {Haiman},\ and\ \citenamefont {Kocsis}}]{Tagawa:2019osr}%
  \BibitemOpen
  \bibfield  {author} {\bibinfo {author} {\bibfnamefont {H.}~\bibnamefont {Tagawa}}, \bibinfo {author} {\bibfnamefont {Z.}~\bibnamefont {Haiman}},\ and\ \bibinfo {author} {\bibfnamefont {B.}~\bibnamefont {Kocsis}},\ }\href {https://doi.org/10.3847/1538-4357/ab9b8c} {\bibfield  {journal} {\bibinfo  {journal} {Astrophys. J.}\ }\textbf {\bibinfo {volume} {898}},\ \bibinfo {pages} {25} (\bibinfo {year} {2020})},\ \Eprint {https://arxiv.org/abs/1912.08218} {arXiv:1912.08218 [astro-ph.GA]} \BibitemShut {NoStop}%
\bibitem [{\citenamefont {Narayan}(2000)}]{Narayan:1999ay}%
  \BibitemOpen
  \bibfield  {author} {\bibinfo {author} {\bibfnamefont {R.}~\bibnamefont {Narayan}},\ }\href {https://doi.org/10.1086/308956} {\bibfield  {journal} {\bibinfo  {journal} {Astrophys. J.}\ }\textbf {\bibinfo {volume} {536}},\ \bibinfo {pages} {663} (\bibinfo {year} {2000})},\ \Eprint {https://arxiv.org/abs/astro-ph/9907328} {arXiv:astro-ph/9907328} \BibitemShut {NoStop}%
\bibitem [{\citenamefont {Edgar}(2004)}]{Edgar:2004mk}%
  \BibitemOpen
  \bibfield  {author} {\bibinfo {author} {\bibfnamefont {R.~G.}\ \bibnamefont {Edgar}},\ }\href {https://doi.org/10.1016/j.newar.2004.06.001} {\bibfield  {journal} {\bibinfo  {journal} {New Astron. Rev.}\ }\textbf {\bibinfo {volume} {48}},\ \bibinfo {pages} {843} (\bibinfo {year} {2004})},\ \Eprint {https://arxiv.org/abs/astro-ph/0406166} {arXiv:astro-ph/0406166} \BibitemShut {NoStop}%
\bibitem [{\citenamefont {Fujita}\ and\ \citenamefont {Hikida}(2009)}]{Fujita:2009bp}%
  \BibitemOpen
  \bibfield  {author} {\bibinfo {author} {\bibfnamefont {R.}~\bibnamefont {Fujita}}\ and\ \bibinfo {author} {\bibfnamefont {W.}~\bibnamefont {Hikida}},\ }\href {https://doi.org/10.1088/0264-9381/26/13/135002} {\bibfield  {journal} {\bibinfo  {journal} {Class. Quant. Grav.}\ }\textbf {\bibinfo {volume} {26}},\ \bibinfo {pages} {135002} (\bibinfo {year} {2009})},\ \Eprint {https://arxiv.org/abs/0906.1420} {arXiv:0906.1420 [gr-qc]} \BibitemShut {NoStop}%
\bibitem [{\citenamefont {Carter}(1968)}]{Carter:1968ks}%
  \BibitemOpen
  \bibfield  {author} {\bibinfo {author} {\bibfnamefont {B.}~\bibnamefont {Carter}},\ }\href {https://doi.org/10.1007/BF03399503} {\bibfield  {journal} {\bibinfo  {journal} {Commun. Math. Phys.}\ }\textbf {\bibinfo {volume} {10}},\ \bibinfo {pages} {280} (\bibinfo {year} {1968})}\BibitemShut {NoStop}%
\bibitem [{\citenamefont {Schmidt}(2002)}]{Schmidt:2002qk}%
  \BibitemOpen
  \bibfield  {author} {\bibinfo {author} {\bibfnamefont {W.}~\bibnamefont {Schmidt}},\ }\href {https://doi.org/10.1088/0264-9381/19/10/314} {\bibfield  {journal} {\bibinfo  {journal} {Class. Quant. Grav.}\ }\textbf {\bibinfo {volume} {19}},\ \bibinfo {pages} {2743} (\bibinfo {year} {2002})},\ \Eprint {https://arxiv.org/abs/gr-qc/0202090} {arXiv:gr-qc/0202090} \BibitemShut {NoStop}%
\bibitem [{\citenamefont {Park}\ and\ \citenamefont {Nasipak}(2024)}]{Park:2024sjj}%
  \BibitemOpen
  \bibfield  {author} {\bibinfo {author} {\bibfnamefont {S.}~\bibnamefont {Park}}\ and\ \bibinfo {author} {\bibfnamefont {Z.}~\bibnamefont {Nasipak}},\ }\href {https://doi.org/10.21105/joss.06587} {\bibfield  {journal} {\bibinfo  {journal} {J. Open Source Softw.}\ }\textbf {\bibinfo {volume} {9}},\ \bibinfo {pages} {6587} (\bibinfo {year} {2024})},\ \Eprint {https://arxiv.org/abs/2406.01413} {arXiv:2406.01413 [gr-qc]} \BibitemShut {NoStop}%
\bibitem [{\citenamefont {{Miniutti}}\ \emph {et~al.}(2019)\citenamefont {{Miniutti}}, \citenamefont {{Saxton}}, \citenamefont {{Giustini}}, \citenamefont {{Alexander}}, \citenamefont {{Fender}}, \citenamefont {{Heywood}}, \citenamefont {{Monageng}}, \citenamefont {{Coriat}}, \citenamefont {{Tzioumis}}, \citenamefont {{Read}}, \citenamefont {{Knigge}}, \citenamefont {{Gandhi}}, \citenamefont {{Pretorius}},\ and\ \citenamefont {{Ag{\'\i}s-Gonz{\'a}lez}}}]{2019Natur.573..381M}%
  \BibitemOpen
  \bibfield  {author} {\bibinfo {author} {\bibfnamefont {G.}~\bibnamefont {{Miniutti}}}, \bibinfo {author} {\bibfnamefont {R.~D.}\ \bibnamefont {{Saxton}}}, \bibinfo {author} {\bibfnamefont {M.}~\bibnamefont {{Giustini}}}, \bibinfo {author} {\bibfnamefont {K.~D.}\ \bibnamefont {{Alexander}}}, \bibinfo {author} {\bibfnamefont {R.~P.}\ \bibnamefont {{Fender}}}, \bibinfo {author} {\bibfnamefont {I.}~\bibnamefont {{Heywood}}}, \bibinfo {author} {\bibfnamefont {I.}~\bibnamefont {{Monageng}}}, \bibinfo {author} {\bibfnamefont {M.}~\bibnamefont {{Coriat}}}, \bibinfo {author} {\bibfnamefont {A.~K.}\ \bibnamefont {{Tzioumis}}}, \bibinfo {author} {\bibfnamefont {A.~M.}\ \bibnamefont {{Read}}}, \bibinfo {author} {\bibfnamefont {C.}~\bibnamefont {{Knigge}}}, \bibinfo {author} {\bibfnamefont {P.}~\bibnamefont {{Gandhi}}}, \bibinfo {author} {\bibfnamefont {M.~L.}\ \bibnamefont {{Pretorius}}},\ and\ \bibinfo {author} {\bibfnamefont {B.}~\bibnamefont {{Ag{\'\i}s-Gonz{\'a}lez}}},\ }\href
  {https://doi.org/10.1038/s41586-019-1556-x} {\bibfield  {journal} {\bibinfo  {journal} {\nat}\ }\textbf {\bibinfo {volume} {573}},\ \bibinfo {pages} {381} (\bibinfo {year} {2019})},\ \Eprint {https://arxiv.org/abs/1909.04693} {arXiv:1909.04693 [astro-ph.HE]} \BibitemShut {NoStop}%
\bibitem [{\citenamefont {Zalamea}\ \emph {et~al.}(2010)\citenamefont {Zalamea}, \citenamefont {Menou},\ and\ \citenamefont {Beloborodov}}]{Zalamea:2010mv}%
  \BibitemOpen
  \bibfield  {author} {\bibinfo {author} {\bibfnamefont {I.}~\bibnamefont {Zalamea}}, \bibinfo {author} {\bibfnamefont {K.}~\bibnamefont {Menou}},\ and\ \bibinfo {author} {\bibfnamefont {A.~M.}\ \bibnamefont {Beloborodov}},\ }\href {https://doi.org/10.1111/j.1745-3933.2010.00930.x} {\bibfield  {journal} {\bibinfo  {journal} {Mon. Not. Roy. Astron. Soc.}\ }\textbf {\bibinfo {volume} {409}},\ \bibinfo {pages} {25} (\bibinfo {year} {2010})},\ \Eprint {https://arxiv.org/abs/1005.3987} {arXiv:1005.3987 [astro-ph.HE]} \BibitemShut {NoStop}%
\bibitem [{\citenamefont {King}(2020)}]{King:2020jtd}%
  \BibitemOpen
  \bibfield  {author} {\bibinfo {author} {\bibfnamefont {A.}~\bibnamefont {King}},\ }\href {https://doi.org/10.1093/mnrasl/slaa020} {\bibfield  {journal} {\bibinfo  {journal} {Mon. Not. Roy. Astron. Soc.}\ }\textbf {\bibinfo {volume} {493}},\ \bibinfo {pages} {L120} (\bibinfo {year} {2020})},\ \Eprint {https://arxiv.org/abs/2002.00970} {arXiv:2002.00970 [astro-ph.HE]} \BibitemShut {NoStop}%
\bibitem [{\citenamefont {Cufari}\ \emph {et~al.}(2022)\citenamefont {Cufari}, \citenamefont {Coughlin},\ and\ \citenamefont {Nixon}}]{Cufari:2022szx}%
  \BibitemOpen
  \bibfield  {author} {\bibinfo {author} {\bibfnamefont {M.}~\bibnamefont {Cufari}}, \bibinfo {author} {\bibfnamefont {E.~R.}\ \bibnamefont {Coughlin}},\ and\ \bibinfo {author} {\bibfnamefont {C.~J.}\ \bibnamefont {Nixon}},\ }\href {https://doi.org/10.3847/2041-8213/ac6021} {\bibfield  {journal} {\bibinfo  {journal} {Astrophys. J. Lett.}\ }\textbf {\bibinfo {volume} {929}},\ \bibinfo {pages} {L20} (\bibinfo {year} {2022})},\ \Eprint {https://arxiv.org/abs/2203.08162} {arXiv:2203.08162 [astro-ph.HE]} \BibitemShut {NoStop}%
\bibitem [{\citenamefont {Nicholl}\ \emph {et~al.}(2024)\citenamefont {Nicholl} \emph {et~al.}}]{Nicholl:2024fff}%
  \BibitemOpen
  \bibfield  {author} {\bibinfo {author} {\bibfnamefont {M.}~\bibnamefont {Nicholl}} \emph {et~al.},\ }\href {https://doi.org/10.1038/s41586-024-08023-6} {\bibfield  {journal} {\bibinfo  {journal} {Nature}\ }\textbf {\bibinfo {volume} {634}},\ \bibinfo {pages} {804} (\bibinfo {year} {2024})},\ \Eprint {https://arxiv.org/abs/2409.02181} {arXiv:2409.02181 [astro-ph.HE]} \BibitemShut {NoStop}%
\bibitem [{\citenamefont {{Sniegowska}}\ \emph {et~al.}(2020)\citenamefont {{Sniegowska}}, \citenamefont {{Czerny}}, \citenamefont {{Bon}},\ and\ \citenamefont {{Bon}}}]{2020A&A...641A.167S}%
  \BibitemOpen
  \bibfield  {author} {\bibinfo {author} {\bibfnamefont {M.}~\bibnamefont {{Sniegowska}}}, \bibinfo {author} {\bibfnamefont {B.}~\bibnamefont {{Czerny}}}, \bibinfo {author} {\bibfnamefont {E.}~\bibnamefont {{Bon}}},\ and\ \bibinfo {author} {\bibfnamefont {N.}~\bibnamefont {{Bon}}},\ }\href {https://doi.org/10.1051/0004-6361/202038575} {\bibfield  {journal} {\bibinfo  {journal} {Astronomy I\& Astrophysics}\ }\textbf {\bibinfo {volume} {641}},\ \bibinfo {eid} {A167} (\bibinfo {year} {2020})},\ \Eprint {https://arxiv.org/abs/2007.06441} {arXiv:2007.06441 [astro-ph.GA]} \BibitemShut {NoStop}%
\bibitem [{\citenamefont {{Raj}}\ and\ \citenamefont {{Nixon}}(2021)}]{2021ApJ...909...82R}%
  \BibitemOpen
  \bibfield  {author} {\bibinfo {author} {\bibfnamefont {A.}~\bibnamefont {{Raj}}}\ and\ \bibinfo {author} {\bibfnamefont {C.~J.}\ \bibnamefont {{Nixon}}},\ }\href {https://doi.org/10.3847/1538-4357/abdc25} {\bibfield  {journal} {\bibinfo  {journal} {\apj}\ }\textbf {\bibinfo {volume} {909}},\ \bibinfo {eid} {82} (\bibinfo {year} {2021})},\ \Eprint {https://arxiv.org/abs/2101.05825} {arXiv:2101.05825 [astro-ph.HE]} \BibitemShut {NoStop}%
\bibitem [{\citenamefont {{Pan}}\ \emph {et~al.}(2022)\citenamefont {{Pan}}, \citenamefont {{Li}}, \citenamefont {{Cao}}, \citenamefont {{Miniutti}},\ and\ \citenamefont {{Gu}}}]{2022ApJ...928L..18P}%
  \BibitemOpen
  \bibfield  {author} {\bibinfo {author} {\bibfnamefont {X.}~\bibnamefont {{Pan}}}, \bibinfo {author} {\bibfnamefont {S.-L.}\ \bibnamefont {{Li}}}, \bibinfo {author} {\bibfnamefont {X.}~\bibnamefont {{Cao}}}, \bibinfo {author} {\bibfnamefont {G.}~\bibnamefont {{Miniutti}}},\ and\ \bibinfo {author} {\bibfnamefont {M.}~\bibnamefont {{Gu}}},\ }\href {https://doi.org/10.3847/2041-8213/ac5faf} {\bibfield  {journal} {\bibinfo  {journal} {The Astrophysical Journal Letters}\ }\textbf {\bibinfo {volume} {928}},\ \bibinfo {eid} {L18} (\bibinfo {year} {2022})},\ \Eprint {https://arxiv.org/abs/2203.12137} {arXiv:2203.12137 [astro-ph.GA]} \BibitemShut {NoStop}%
\bibitem [{\citenamefont {{Syer}}\ \emph {et~al.}(1991)\citenamefont {{Syer}}, \citenamefont {{Clarke}},\ and\ \citenamefont {{Rees}}}]{1991MNRAS.250..505S}%
  \BibitemOpen
  \bibfield  {author} {\bibinfo {author} {\bibfnamefont {D.}~\bibnamefont {{Syer}}}, \bibinfo {author} {\bibfnamefont {C.~J.}\ \bibnamefont {{Clarke}}},\ and\ \bibinfo {author} {\bibfnamefont {M.~J.}\ \bibnamefont {{Rees}}},\ }\href {https://doi.org/10.1093/mnras/250.3.505} {\bibfield  {journal} {\bibinfo  {journal} {Monthly Notices of the Royal Astronomical Society}\ }\textbf {\bibinfo {volume} {250}},\ \bibinfo {pages} {505} (\bibinfo {year} {1991})}\BibitemShut {NoStop}%
\bibitem [{\citenamefont {Peters}(1964)}]{Peters:1964zz}%
  \BibitemOpen
  \bibfield  {author} {\bibinfo {author} {\bibfnamefont {P.~C.}\ \bibnamefont {Peters}},\ }\href {https://doi.org/10.1103/PhysRev.136.B1224} {\bibfield  {journal} {\bibinfo  {journal} {Phys. Rev.}\ }\textbf {\bibinfo {volume} {136}},\ \bibinfo {pages} {B1224} (\bibinfo {year} {1964})}\BibitemShut {NoStop}%
\bibitem [{\citenamefont {Mandel}(2021)}]{Mandel:2021fra}%
  \BibitemOpen
  \bibfield  {author} {\bibinfo {author} {\bibfnamefont {I.}~\bibnamefont {Mandel}},\ }\href {https://doi.org/10.3847/2515-5172/ac2d35} {\bibfield  {journal} {\bibinfo  {journal} {Res. Notes AAS}\ }\textbf {\bibinfo {volume} {5}},\ \bibinfo {pages} {223} (\bibinfo {year} {2021})},\ \Eprint {https://arxiv.org/abs/2110.09254} {arXiv:2110.09254 [astro-ph.HE]} \BibitemShut {NoStop}%
\end{thebibliography}%

\end{document}